\documentclass{article}
\usepackage{soul}
\usepackage{graphicx, graphics}
\usepackage{syntonly}
\usepackage{bm}
\usepackage{amssymb, amsmath, amsfonts}
\usepackage{subfigure}
\usepackage{multirow}
\usepackage{color}
\usepackage{hyperref}

\newcommand{\bx}{\mathbf X}
\newcommand{\bu}{\mathbf u}

\newcommand{\by}{\bm x}
\newcommand{\bF}{\mathbf F}

\newcommand{\Grad}{\mbox{Grad}}
\newcommand{\tnabla}{\tilde{\nabla}}
\newcommand{\bT}{\mathbf T}
\newcommand{\bK}{\mathbf L}
\newcommand{\bP}{\mathbf P}

\newcommand{\bY}{\bm x_{ij}}

\newcommand{\ud}{\mathrm d}
\newcommand{\bB}{\mathbf \Omega}

\newcommand{\etal}{\textit{et al }}

\newcommand{\fracp}[2]{\frac{\partial#1 }{\partial #2}}

\newcommand{\sfrac}[2]{#1/#2}

\newcommand{\br}{\bm r}

\newcommand{\cS}{\mathcal S}

\newcommand{\tr}[1]{\mathrm{tr}({#1})}

\renewcommand{\hl}{\textcolor{black}}
\newcommand{\tdelta}{\bar{\delta}}
\newcommand{\fF}{\mathcal F}

\begin{document}
\title{Dual-support smoothed particle hydrodynamics in solid: variational principle and implicit formulation}
\author{Huilong Ren,Xiaoying Zhuang,\\
Timon Rabczuk,HeHua Zhu\\
Division of Computational Mechanics, \\
Ton Duc Thang University, Ho Chi Minh City, Viet Nam\\
Faculty of Civil Engineering,\\
 Ton Duc Thang University, Ho Chi Minh City, Viet Nam\\
Institute of Structural Mechanics, \\
Bauhaus-University Weimar, 99423 Weimar, Germany\\
 Institute of Conitnuum Mechanics, \\
 Leibniz University Hannover, Hannover, Germany\\
 State Key Laboratory of Disaster Reduction in Civil Engineering,\\
  College of Civil Engineering,Tongji University, \\
  Shanghai 200092, China}
\maketitle

\begin{abstract}
We derive the dual-support smoothed particle hydrodynamics (DS-SPH) in solid {within} the framework of variational principle. The tangent stiffness matrix of SPH {can be} obtained with ease, {and} can be served as the basis for {the present} implicit SPH. We propose an hourglass energy functional, which allows the direct derivation of hourglass force and hourglass tangent stiffness matrix. The dual-support is {involved} in all derivations based on variational principles and is automatically satisfied in the assembling of stiffness matrix. The implementation of stiffness matrix comprises with two steps, the nodal assembly based on deformation gradient and global assembly on all nodes. Several numerical examples are presented to validate the method.
\end{abstract}

%
\section{Introduction}\label{sec:introduction}
Smoothed particle hydrodynamics (SPH) was introduced by Lucy \cite{lucy1977numerical} and Gingold and Monaghan \cite{gingold1977smoothed} to solve astrophysical problems such as the formation of stars and the evolution of dust clouds. Due to its flexibility, SPH has been extended to solve various engineering problems, i.e. free-surface flowing \cite{violeau2016smoothed}, metal cutting \cite{limido2007sph}, impacting simulation \cite{zhou2007three,liu2006adaptive}, brittle/ductile fractures \cite{batra2004analysis}, plate and shell \cite{maurel2008sph,caleyron2012dynamic}, for more complete review of SPH, we refer to \cite{monaghan2005smoothed,liu2010smoothed}. One of the key feature of SPH is that the kernel approximation can convert the PDEs into simple algebraic equations, on which the solutions of the underlying PDEs are obtained. \hl{In contrast with finite element methods \cite{zienkiewicz1977finite,bonet1997nonlinear} and boundary element methods \cite{Fu2018Aug,fu2018singular}}, SPH method discretizes the continuous domain into a set of particles, each particle is associated with physical quantities such as mass, internal energy and velocity. Since no mesh is required, the SPH is considered as one of the oldest meshless methods. Though some advantages over finite element method (FEM) in arbitrarily large deformations and discontinuity modeling such as fractures, SPH is less accurate and robust than mesh-based methods due to the tensile instabilities and rank-deficiency in the nodal integration approach. A number of different schemes are devised to enhance the stability of SPH, such as artificial viscosity \cite{monaghan1992smoothed}, XSPH time integration scheme \cite{monaghan1989problem}, stress points method \cite{dyka1995approach,randles1996smoothed} for rank-deficiency problem, Lagrange kernel \cite{Rabczuk2004} for tensile instabilities, hourglass force method for zero-energy mode \cite{ganzenmuller2015hourglass}. Meanwhile, various techniques have been developed through the years to alleviate these problems, among which include  Corrected Smoothed Particle Method (CSPM) \cite{chen1999completeness}, Reproducing Kernel Particle Method (RKPM) \cite{LiuJunZhang1995}, Symmetric Smoothed Particle Hydrodynamics (SSPH) \cite{zhang2009symmetric}, Optimal Transportation Meshless method (OTM) \cite{Li2010} and so on. 


Related to the variational derivation of SPH, Bonet and Lok \cite{Bonet1999} derived the governing equations of SPH for fluid under the condition of constant smoothing length. Grenier \etal \cite{grenier2009hamiltonian} derived an Hamiltonian interface SPH formulation for multi-fluid
and free surface flows. Price and Monoghan \cite{price2004smoothed} presented variational derivation of Smoothed Particle Hydrodynamics and Magnetohydrodynamics. However, these derivations are limited to fluid. In the spirit of dual-horizon peridynamics \cite{Ren2015,ren2017dual} which is proposed for the purpose of computational efficiency and variable smoothing lengths, we derive the dual-support SPH in solid by variational principle.

The purpose of this paper is to derive by variational principles the dual-support SPH and furthermore construct the tangent stiffness matrix for implicit analysis without zero-energy mode. There are primarily three innovations in the paper. Firstly, we find a direct and simple way to construct the tangent stiffness matrix of SPH in solid. With tangent stiffness matrix, a lot of implicit solvers can be used to find the solution. Secondly, we established a hourglass energy functional and found a simple hourglass force to suppress the hourglass mode in SPH solid. The hourglass force is derived from the requirement of linear completeness, which is different with the stress point scheme and the least-squares stabilization scheme \cite{beissel1996nodal}. The tangent stiffness matrix of hourglass energy can be constructed with ease. Last but not the least, we proposed a framework for the implementation of implicit SPH where the material nonlinearity and geometrical nonlinearity can be included. 

The content of the paper is outlined as follows. In section 2, we review the basic concepts of support and dual-support and derive the dual-support SPH based on variational principles. In order to remove the hourglass mode, we introduce the hourglass energy functional, based on which the hourglass force, the hourglass residual and tangent stiffness matrix are derived in section 3. The implementation and material constitutions are provided in section 4. With the aid of the variation of the deformation gradient tensor, the nodal tangent stiffness matrix is simply the matrix multiplication of common terms. In order to verify the implicit scheme, we give in section 5 four numerical examples in 2D/3D. The numerical results are compared with the theoretical solutions or finite element results {as reference solutions} and the good agreements are obtained. The performance of hourglass control are also analyzed in the same section. Finally, conclusions of the present work are given in section 6.

\section{Variational derivation of dual-support SPH}
\begin{figure}[htp]
 \centering
 \subfigure[]{
 \label{fig:Coord}
 \includegraphics[width=.4\textwidth]{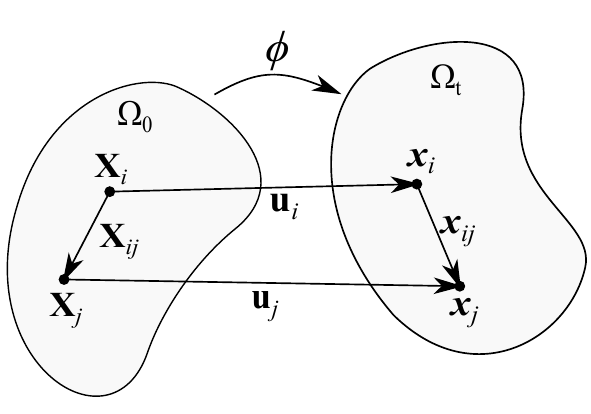}}
 \vspace{.1in}
 \subfigure[]{
 \label{fig:4support}
 \includegraphics[width=.55\textwidth]{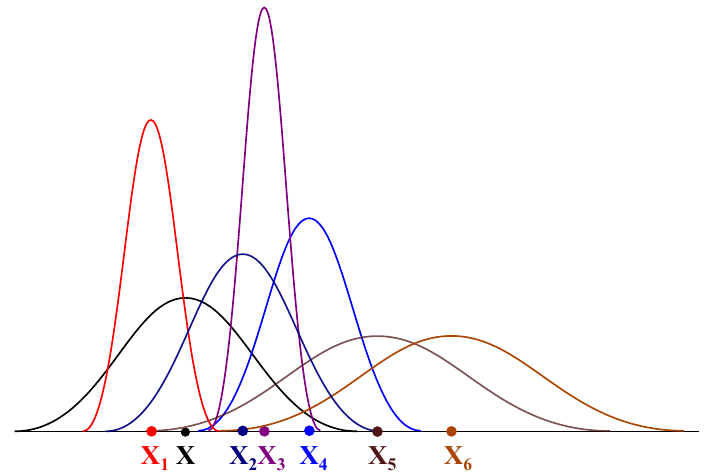}}\\
 \vspace{.3in}
\caption{(a). Configuration for deformed body.(b) Schematic diagram for support and dual-support in one dimension with cubic kernel function. $\cS_{\bx}=\{\bx_1,\bx_2,\bx_3,\bx_4\} $, $\cS_\bx'=\{\bx_1,\bx_2,\bx_5\}$.}
\end{figure} 

Consider a solid in the initial and current configuration as shown in Fig. \ref{fig:Coord}. Let $\bx_i$ be material coordinates in the initial configuration $\mathbf \Omega_0$. A function $\phi$ mapping any point $\bx$ in the reference coordinates to the current coordinate $\by$ at time $t$, 
\begin{align}
\by=\phi(\bx,t).
\end{align}
Let $\by_i:=\phi(\bx_i,t)$ and $\by_j:=\phi(\bx_j,t)$ be the spatial coordinates in the current configuration $\mathbf \Omega_t$ of the corresponding particles; $\bx_{ij}:=\bx_j-\bx_i$ is initial spatial vector, the relative distance vector between $\bx_i$ and $\bx_j$; $\bu_i:=\by_i-\bx_i$ and $\bu_j:=\by_j-\bx_j$ are the displacement vectors for $\bx_i$ and $\bx_j$, respectively; $\bu_{ij}:=\bu_j-\bu_i$ is the relative displacement vector for spatial vector $\bx_{ij}$; $\bY:=\phi(\bx_j,t)-\phi(\bx_i,t)=\bx_{ij}+\bu_{ij}$ is the current spatial vector for $\bx_{ij}$.

The governing equations for SPH solid in Lagrangian description include 
\begin{align}
\rho_0&=\rho\det (\bF) \\
\rho_0 \ddot{\by}&=\nabla_\bx\cdot \bP+\mathbf f\\
\rho_0 \dot{e}&=\bP:\dot{\bF},
\end{align}
where $\bF$ is the deformation gradient, $\bP$ is the first Piola-Kirchhoff stress, $e$ is the internal energy density. In the case of pure elastic solid, the continuity equation and the energy equation can be ignored and only the equation of motion is required.


\textbf{Support} $\cS_i$ is the domain where any particle $\bx_j$ with $X_{ij}=|\bx_{ij}|\leq h_i$, where $h_i$ is the smoothing length for particle $i$. The support $\cS_i $ is usually presented by a spherical domain with radius of $h_i$,
\begin{align}
\cS_{i}=\{\bx_j|X_{ij}\leq h_i\}.
\end{align}
\textbf{Dual-support} is defined as a union of the points whose supports include $\bx_i$, denoted by 
\begin{align}
\cS_i'=\{\bx_j|\bx_i \in \cS_{j}\}=\{\bx_j|X_{ij} \leq h_j\}. \label{eq:dualsupport}
\end{align}
One example to illustrate the support and dual-support is shown in Fig.\ref{fig:4support}.

SPH approximation for a scalar function in the reference of material configuration can be written as
\begin{align}
f(\bx_i)=\sum_{\cS_i} V_j f(\bx_j)W_i(\bx_{ij}),
\end{align}
where $W_i(\bx_{ij})$ is the SPH kernel function for particle $\bx_i$, which only depends on the distance vector between $\bx_i$ and $\bx_j$. $V_j$ is the volume associated with particle $\bx_j$ in the initial configuration. 

The symmetric SPH approximation of a derivative of scalar function $f$ is obtained by the gradient operator on the kernel function,
\begin{align}
\Grad(f(\bx_i))=\sum_{\cS_i} V_j \big(f(\bx_j)-f(\bx_i)\big)\nabla W_i(\bx_{ij}),
\end{align}
where Grad denotes the gradient operator based on the initial configuration, the gradient of the kernel function is calculated by
\begin{align}
\nabla W_i(\bx_{ij})=\frac{\ud W_i(\bx_{ij})}{\ud X_{ij}} \frac{\bx_{ij}}{X_{ij}}.
\end{align}
For the condition of zeroth-order and first-order completeness, the corrected kernel gradient is defined as
\begin{align}
\tnabla W_i(\bx_{ij})=\bK_i^{-1} \nabla W_i(\bx_{ij}),
\end{align}
where the correction matrix $\bK_i$ is defined as
\begin{align}
\bK_i=\sum_{\cS_i} V_j \nabla W_i(\bx_{ij})\otimes \bx_{ij}.
\end{align}
The deformation gradient $\mathbf{F}$ for $\bx_i$ in SPH is defined as 
\begin{align}
\mathbf{F}_{i}=\frac{\partial \bm x_i}{\partial \bx_i}=\sum_{\cS_i}\by_{ij} \otimes \tnabla W_i(\bx_{ij}) V_j. \label{eq:Fdef}
\end{align}
The variation of the deformation gradient 
\begin{align}
\delta \mathbf{F}_{i}=\sum_{\cS_i}\delta\by_{ij} \otimes \tnabla W_i(\bx_{ij}) V_j=\sum_{\cS_i}(\delta\by_j-\delta\by_i) \otimes \tnabla W_i(\bx_{ij}) V_j.
\end{align}
{Let} $\fF(\bF_i)$ {be the strain energy density functional that only depends on the deformation gradient}.
The {first-order} variation of strain energy $\fF(\bF_i)$ in $\cS_i$ of point $\bx_i$ is 
\begin{align}
\delta \fF(\bF_i)=\fracp{\fF}{\bF_i}\cdot \delta\bF_i=\fracp{\fF}{\bF_i}\cdot \sum_{\cS_i}(\delta\by_j-\delta\by_i) \otimes \tnabla W_i(\bx_{ij}) V_j\notag\\
=\bP_i\cdot\sum_{\cS_i}(\delta\by_j-\delta\by_i) \otimes \tnabla W_i(\bx_{ij}) V_j.
\end{align}
The first Piola-Kirchhoff stress $\mathbf P$ related to the deformation gradient is {then} given by
\begin{align}
\mathbf P=\fracp{\fF(\mathbf{F})}{\mathbf{F}}\label{eq:pf}.
\end{align}
The Lagrangian for the system includes the kinetic energy, potential energy (strain energy, the body force energy and external work), and can be expressed as
\begin{align}
L(\dot{\by},\by)=\sum_{V_i\in \bB_0}\Big(\frac{1}{2}\rho \dot{\by}_i\cdot \dot{\by}_i-\fF(\bF_i)+\mathbf b_0 \cdot (\by_i-\bx_i)\Big) V_i+\int_{\Gamma_0} \mathbf f_0 \cdot (\by-\bx)\mathrm{d}\Gamma_0.
\end{align}
The external work in time interval $[t_1,t_2]$ is $W^{ext}=\int_{t_1}^{t_2} \int_{\Gamma_0} \mathbf f_0 \cdot (\by-\bx)\mathrm{d}\Gamma_0\mathrm{d}t$. 
The integral of the Lagrangian $L$ between two instants of time $t_1$ and $t_2$ is $S=\int_{t_1}^{t_2} L(\dot{\by},\by)\mathrm{d}t$. {In order to derive the internal force between particles, we neglect the external work for simplicity}. Applying the principle of least action, we have
\begin{align}
&\delta S=\int_{t_1}^{t_2}\sum_{V_i\in \bB_0}\Big(\rho \dot{\by}_i\cdot \delta\dot{\by}_i-\delta \fF(\bF_i)+\mathbf b_0 \cdot \delta (\by_i-\bx_i)\Big) V_i\mathrm{d}t\notag\\
&=\int_{t_1}^{t_2}\sum_{V_i\in \bB_0}\Big(-\rho \ddot{\by}_i\cdot \delta\by_i-\sum_{\cS_i} \bP_i\cdot (\delta \by_j-\delta \by_i) \otimes \tnabla W_i(\bx_{ij}) V_j +\mathbf b_0 \cdot \delta \by_i\Big) V_i\mathrm{d}t\notag\\
&=\int_{t_1}^{t_2}\sum_{V_i\in \bB_0}\Big(\bigl(-\rho \ddot{\by}_i+\sum_{\cS_i} \bP_i\cdot \tnabla W_i(\bx_{ij}) V_j-\sum_{\cS_i'} \bP_j \cdot \tnabla W_j(\bx_{ji}) V_j+\mathbf b_0\big)\cdot \delta{\by_i}\Big)V_{i}\mathrm{d}t.\label{eq:deltaS}
\end{align}
The derivation considers the boundary condition $\delta \by(t_1)=0,\,\delta \by(t_2)=0$. In the second and third step, the dual-support is considered as follows. In the second step, the term with $\delta \by_j$ is the force vector from $\bx_i$'s support, but is added to particle $\bx_j$; since $\bx_j\in \cS_i$, $\bx_i$ belongs to the dual-support $\cS'_{j}$ of $\bx_j$. In the third step, all terms with $\delta \by_i$ are collected from other particles whose supports contain $\bx_i$ and therefore form the dual-support of $\bx_i$. For any $\delta \by_i$, the first order variation $\delta S=0$ {leading to}
\begin{align}
\rho \ddot{\by}_{i}=\sum_{\cS_i} \bP_i \cdot \tnabla W_i(\bx_{ij}) V_j-\sum_{\cS'_{i}} \bP_j \cdot \tnabla W_j(\bx_{ji}) V_j+\mathbf b_0, \forall \bx_i \in \Omega_0. \label{eq:gsph}
\end{align}

In the paper, we use the kernel function of quintic spline
\begin{align}
W(\br)=\alpha_d \Big((1-r/h)_+^5-6(\frac{2}{3}-r/h)_+^5+15(\frac{1}{3}-r/h)_+^5\Big)
\end{align}
where $r=\|\br\|$, $h$ is the smoothing length scale; $\alpha_d= (\sfrac{3^5}{40},\, {\sfrac{3^7 7}{478 \pi}},\,{\sfrac{3^7}{40\pi}})$ for 1,2,3 dimensional spaces, respectively; $x_+=\mbox{max}(0,x)$. For more kernel functions \hl{with discussions on their properties,} we refer to \cite{dehnen2012improving}. \hl{Based on our numerical test, the selection of kernel functions has very small influence on the final result.}

\section{Functional of hourglass energy}
In order to remove the hourglass mode (zero-energy mode), the conventional SPH adds a penalty term to the force state, in which the penalty force is proportional to the difference between current location of a point and the position predicted by the deformation gradient \cite{ganzenmuller2015hourglass}:
\begin{align}
\mathbf T^{hg}_{i}\propto \sum_{\cS_i} \frac{(\bF_i \bx_{ij}-\by_{ij})\cdot \by_{ij}}{x_{ij}}.
\end{align}
However, the above formulation is only feasible in the explicit formulation since $x_{ij}$ exists in the denominator.

The displacement field in the neighborhood of a particle is required to be linear. Therefore, it has to be exactly described by the deformation gradient, and the hourglass modes are identified as that part of the displacement field, which is not described by the deformation gradient \cite{ganzenmuller2015hourglass}. In practice, the difference of current deformed vector $\bY$ and predicted vector by deformation gradient is $(\bF_i\bx_{ij}-\bY)$. We formulate the hourglass energy based on the difference in the support as follows. Let $\alpha=\frac{\mu}{m_{\bK_i}}$ be a coefficient for the hourglass energy, where ${m_{\bK_i}}=\tr{\bK}$, $\mu$ is the shear modulus, the functional for zero-energy mode is defined as
\begin{align}
\fF^{hg}_i&=\alpha \sum_{\cS_i} \frac{1}{X_{ij}}\frac{\ud W_i(\bx_{ij})}{\ud X_{ij}} (\bF_i\bx_{ij}-\bY)^T (\bF_i\bx_{ij}-\bY) V_j \notag\\
&=\alpha \sum_{\cS_i} \frac{1}{X_{ij}}\frac{\ud W_i(\bx_{ij})}{\ud X_{ij}}\Big(\bx_{ij}^T \bF_i^T \bF_i\bx_{ij}+\bY^T \bY-2 \bY^T \bF_i\bx_{ij}\Big) V_j \notag\\
&=\alpha \sum_{\cS_i} \frac{1}{X_{ij}}\frac{\ud W_i(\bx_{ij})}{\ud X_{ij}}\Big(\bF_i^T \bF_i: \bx_{ij} \otimes \bx_{ij}+\bY^T \bY-2 \bF_i:\bY\otimes\bx_{ij}\Big) V_j \notag\\
&=\alpha \bF_i^T \bF_i: \sum_{\cS_i} \frac{1}{X_{ij}}\frac{\ud W_i(\bx_{ij})}{\ud X_{ij}}\bx_{ij} \otimes \bx_{ij} V_j +\alpha \sum_{\cS_i} \frac{1}{X_{ij}}\frac{\ud W_i(\bx_{ij})}{\ud X_{ij}} \bY^T \bY V_j \notag\\&\quad-2 \alpha \bF_i:\sum_{\cS_i} \frac{1}{X_{ij}}\frac{\ud W_i(\bx_{ij})}{\ud X_{ij}}\bY\otimes\bx_{ij} V_j \notag\\
&=\alpha \bF_i^T \bF_i: \sum_{\cS_i} \bx_{ij} \otimes \nabla W_i(\bx_{ij}) V_j +\alpha \sum_{\cS_i} \frac{\ud W_i(\bx_{ij})}{\ud X_{ij}} \frac{\bY^T}{X_{ij}} \bY V_j -2 \alpha \bF_i:\sum_{\cS_i} \bY\otimes\nabla W_i(\bx_{ij}) V_j \notag\\
&=\alpha \bF_i^T \bF_i: \bK_i+\alpha \sum_{\cS_i} \frac{1}{X_{ij}}\frac{\ud W_i(\bx_{ij})}{\ud X_{ij}} \bY^T \bY V_j -2 \alpha \bF_i: (\bF_i \bK_i)\notag\\
&=\frac{\mu}{m_{\bK_i}} \Big(\sum_{\cS_i} \frac{\ud W_i(\bx_{ij})}{\ud X_{ij}}\frac{\bY^T \bY}{X_{ij}} V_j - \bF_i: (\bF_i\bK_i)\Big).
\end{align}
The above definition of hourglass energy is similar to the variance in probability theory and statistics. In above derivation, we used the relations:
$ \bF^T \bF: \bK=\bF: (\bF \bK), \mathbf a^T \mathbf M \mathbf b=\mathbf M: \mathbf a\otimes \mathbf b, \mathbf A:\mathbf B=\tr{\mathbf A \mathbf B^T}$, where capital letter denotes matrix and small letter is column vector. The purpose of $m_{\bK}$ is to make the energy functional independent with the support since $\bK$ is involved in $\bF^T \bF: \bK$.

In order to derive the residual and tangent stiffness matrix directly, some notation to denote the variation and how the variations are related to the residual and stiffness matrix are introduced subsequently. Assume a functional $\fF(\bu)$, where $\bu$ are unknown function vector, the first and second variations can be expressed as
\begin{align}
\delta \fF(\bu)&=\partial_{\bu} \fF(\bu) \cdot\delta \bu=\tdelta \fF(\bu)\cdot \delta \bu\\
\delta^2 \fF(\bu)&=\partial_{\bu\bu} \fF(\bu)\cdot \delta \bu\delta \bu=\tdelta^2 \fF(\bu) \cdot \delta \bu\delta \bu\notag
\end{align}
where the special variation $\tdelta \fF(\bu)$ and $\tdelta^2 \fF(\bu)$ are defined as
\begin{align}
\tdelta \fF(\bu)&:=\partial_{\bu} \fF(\bu)\\
\tdelta^2 \fF(\bu)&:=\partial_{\bu\bu} \fF(\bu)
\end{align}
The gradient vector and Hessian matrix represent the residual vector and tangent stiffness matrix of the functional, respectively, with unknown functions $\bu$ being the independent variables. Hence,
\begin{align}
&\mathbf R(\bu)=\tdelta \fF(\bu)\notag\\
&\mathbf K(\bu)=\tdelta^2 \fF(\bu)\notag
\end{align}

For example, when $\bu=[u,v]$, the special variations of functional $\fF(u,v)$ are given as
\begin{align}
\tdelta \fF(u,v)&=\partial_u \fF \tdelta u+\partial_v \fF\tdelta v=[\partial_u \fF,\partial_v \fF]\notag\\
\tdelta^2 \fF(u,v)&=\partial_{uu} \fF \tdelta u\tdelta u+\partial_{uv} \fF \tdelta u\tdelta v+\partial_{vu} \fF\tdelta v \tdelta u+\partial_{vv} \fF\tdelta v \tdelta v\notag\\
&=\begin{bmatrix}\partial_{uu} \fF & \partial_{uv} \fF\\ \partial_{vu} \fF& \partial_{vv} \fF\end{bmatrix}\notag
\end{align}
where $\tdelta u$ has no other meaning but denotes the index of $\partial_u \fF$ in residual vector by the index of $u$ in the unknown vector. Namely, the term $\partial_v \fF\tdelta v$ represents $\partial_v \fF$ be in the second location of the residual vector since $v$ is in the second position of $[u,v]$. The term $\partial_{uv} \fF \tdelta u\tdelta v$ denotes that the location of $\partial_{uv} \fF$ is (1,2), while the term $\partial_{vu} \fF \tdelta v\tdelta u$ denotes that the location of $\partial_{vu} \fF$ is (2,1). The special first-order and second-order variations of a functional lead to the residual and tangent stiffness matrix directly. The traditional variation can be recovered by the inner product of the special variation and the variation of the unknown vector. 

Therefore, the variation of $\bF_i: \bF_i\bK_i$ can be rewritten as
\begin{align}
\tdelta(\bF_i: \bF_i\bK_i)&=\tdelta(\bF_i\bK_i:\bF_i)=2 \bF_i\bK_i:\tdelta \bF_i\notag\\
&=2 \bF_i\bK_i:\sum_{\cS_i}  \tdelta \by_{ij}\otimes \tnabla W_i(\bx_{ij})V_j\notag\\
&=2 \sum_{\cS_i}  \tdelta \by_{ij}^T \bF_i\bK_i \tnabla W_i(\bx_{ij})V_j\notag\\
&=2 \sum_{\cS_i}  \tdelta \by_{ij}^T (\bF_i \nabla W_i(\bx_{ij}))V_j\notag\\
&=2 \sum_{\cS_i}  (\bF_i \nabla W_i(\bx_{ij}))\cdot \tdelta\by_{ij}V_j.
\end{align}
Then the variation of $\fF^{hg}$ is
\begin{align}
\mathbf R^{hg}_i&=\tdelta \fF^{hg}_i\notag\\
&=\frac{\mu}{ m_{\bK_i}} \Big(\sum_{\cS_i} \frac{1}{X_{ij}}\frac{d W_i(\bx_{ij})}{d X_{ij}}\tdelta( \by_{ij}\cdot  \by_{ij})  V_j-\tdelta( \bF_i: \bF_i\bK_i)\Big)\notag\\
&=\frac{\mu}{ m_{\bK_i}} \Big(\sum_{\cS_i}2 \frac{1}{X_{ij}}\frac{d W_i(\bx_{ij})}{d X_{ij}} \by_{ij}\cdot  \tdelta\by_{ij} V_j-2 \sum_{\cS_i}  (\bF_i \nabla W_i(\bx_{ij}))\cdot \tdelta \by_{ij}V_j\Big)\notag\\
&=\frac{2\mu}{ m_{\bK_i}} \sum_{\cS_i} \frac{1}{X_{ij}}\frac{d W_i(\bx_{ij})}{d X_{ij}} (\by_{ij}-\bF_i \bx_{ij})\cdot (\tdelta \by_j-\tdelta \by_i) V_j\label{eq:Phg}
\end{align}
$\mathbf R^{hg}_i$ is the residual for hourglass energy. 
 Eq.\ref{eq:Phg} gives the explicit formula for the hourglass force. The term on $\tdelta \by_i$ is the hourglass force from its support, while the terms on $\tdelta \bu'$ are the hourglass forces for the dual support $\cS_{j}'$ of point $\bx_j$. When the displacement field is consistent with the deformation gradient, then the hourglass energy residual is zero. For individual vector $\bx_{ij}$, the hourglass force vector can be obtained the same way as Eq.\ref{eq:deltaS}, 
\begin{align}
\bT_{{ij}}^{hg}=-\Big(\delta \fF^{hg}\Big)_{ij}=-\frac{2\mu}{m_{\bK_i}} \frac{1}{X_{ij}}\frac{\ud W_i(\bx_{ij})}{\ud X_{ij}}\big(\bY-\bF_{i}\bx_{ij}\big)=\frac{2\mu}{m_{\bK_i}} \big(\bF_{i}\nabla W_i(\bx_{ij})-\frac{\bY}{X_{ij}}\frac{\ud W_i(\bx_{ij})}{\ud X_{ij}}\big).
\end{align}

The governing equation with hourglass force is
\begin{align}
\rho \ddot{\bu}_{i}=\sum_{\cS_i} \Big(\mathbf{P}_{i} \cdot \tnabla W_i(\bx_{ij}) +\bT_{{ij}}^{hg}\Big)V_j-\sum_{\cS'_{i}}\Big(\mathbf{P}_{j}\cdot \tnabla W_j(\bx_{ji})+\bT_{ji}^{hg}\Big) V_j+\mathbf b_0. \label{eq:pdhg}
\end{align}
\hl{One can see from Eq.\ref{eq:pdhg} that one's particle's hourglass forces are divided into two groups, these from the support and the others from the dual-support. The derivation of the hourglass force is similar to the variational derivation on strain energy functional, thus is consistent with dual-support configuration. The hourglass forces from the support can be viewed as the direct forces, while the hourglass forces from the dual-support are the reaction forces. Therefore, the hourglass forces follows the Newton's third law, the same as internal forces. When variable smoothing lengths are used, the hourglass forces have no influence on the conservations of linear momentum and angular momentum}.

The variation of $\tdelta \fF^{hg}$ leads to the hourglass tangent stiffness matrix,
\begin{align}
\mathbf K^{hg}_i=\tdelta^2 \fF^{hg}_i=\frac{\mu}{m_{\bK_i}} \Big(\sum_{\cS_i} \frac{\ud W_i(\bx_{ij})}{\ud X_{ij}}\frac{1}{X_{ij}}(\tdelta \by_j-\tdelta \by_i)^T (\tdelta \by_j-\tdelta \by_i) V_j- \tdelta\bF_i \bK_i:\tdelta\bF_i \Big).\label{eq:Khg}
\end{align}

Similarly, the hourglass correction for scalar field is
\begin{align}
T_{{ij}}^{hg}=-\Big(\tdelta \fF^{hg}\Big)_{ij}=\frac{2\mu}{m_{\bK_i}} \big(\nabla s_{i}\cdot\nabla W_i(\bx_{ij})-\frac{s_{ij}}{X_{ij}}\frac{\ud W_i(\bx_{ij})}{\ud X_{ij}}\big).
\end{align}
where $s_{ij}=s_j-s_i$.
\vspace{-2pt}
\section{\hl{Numerical implementation}}
For elastic material, the strain energy density is a function of the deformation gradient. For the total Lagrange formulation, it is convenient to use the first Piola-Kirchhoff stress, which is the direct derivative of the strain energy over the deformation gradient,
\begin{align}
\mathbf P=\frac{\partial \psi(\bF)}{\partial \bF},
\end{align}
where
\begin{align}
\bF=\begin{bmatrix}F_1 & F_2 & F_3\\F_4& F_5 & F_6\\F_7& F_8 & F_9\end{bmatrix}
\end{align}
Furthermore, the material tensor (stress-strain relation) which is required in the implicit analysis can be obtained with the derivative of the first Piola-Kirchhoff stress,
\begin{align}
\mathbf D_4=\frac{\partial \mathbf P}{\partial \mathbf F}=\fracp{^2 \psi(\bF)}{\bF^T \partial \bF}.
\end{align}
The 4th order material tensor $\mathbf D_4$ can be expressed in matrix form when the deformation gradient is flattened.
\begin{align}
{\mathbf D}=\begin{bmatrix}
\frac {\partial P_1}{\partial F_{1}} & \frac {\partial P_1}{\partial F_{2}}&\cdots &\frac {\partial P_1}{\partial F_{9}}\\
\frac {\partial P_2}{\partial F_{1}} & \frac {\partial P_2}{\partial F_{2}}&\cdots &\frac {\partial P_2}{\partial F_{9}}\\
\vdots &\vdots &\ddots &\vdots \\
\frac {\partial P_9}{\partial F_{1}} & \frac {\partial P_1}{\partial F_{2}}&\cdots &\frac {\partial P_9}{\partial F_{9}}\\
\end{bmatrix}
=
\begin{bmatrix}
\frac {\partial ^{2}\psi(F)}{\partial F_{1}^{2}} & \frac {\partial ^{2}\psi(F)}{\partial F_{1}\,\partial F_{2}}&\cdots &\frac {\partial^{2}\psi(F)}{\partial F_{1}\,\partial F_{9}}\\
\frac {\partial ^{2}\psi(F)}{\partial F_{2}\partial F_{1}} & \frac {\partial ^{2}\psi(F)}{\partial F_{2}\,\partial F_{2}}&\cdots &\frac {\partial^{2}\psi(F)}{\partial F_{2}\,\partial F_{9}}\\
\vdots &\vdots &\ddots &\vdots \\
\frac {\partial ^{2}\psi(F)}{\partial F_{9}\partial F_{1}} & \frac {\partial ^{2}\psi(F)}{\partial F_{9}\,\partial F_{2}}&\cdots &\frac {\partial^{2}\psi(F)}{\partial F_{9}\,\partial F_{9}}\\
\end{bmatrix},
\end{align}
where the flattened deformation gradient and first Piola-Kirchhoff stress are
\begin{align}
F=(F_1, F_2, F_3,F_4,F_5, F_6,F_7,F_8,F_9),
\end{align}
and 
\begin{align}
P=\frac{\partial \psi(F)}{\partial F}=(\frac{\partial \psi(F)}{\partial F_1},\frac{\partial \psi(F)}{\partial F_2},\cdots,\frac{\partial \psi(F)}{\partial F_9}).
\end{align}
The derivative of the determinant of deformation gradient on $\bF$ is
\begin{align}
J=\det(\bF), J_{,\bF}=\begin{bmatrix}F_5 F_9-F_6 F_8 & F_6 F_7-F_4 F_9 & F_4 F_8-F_5 F_7\\F_3 F_8-F_2 F_9& F_1 F_9-F_3 F_7 & F_2 F_7-F_1 F_8\\F_2 F_6-F_3 F_5& F_3 F_4-F_1 F_6 & F_1 F_5-F_2 F_4\end{bmatrix}.
\end{align}
Since the strain energy is formulated on the particles, the total discrete strain energy is the sum of all strain energy on the particles,
\begin{align}
\fF=\sum_{i=1}^N V_i \psi(F_i),
\end{align}
where $V_i$ is the volume associated to particle $i$, $N$ is the number of particles, $F_i$ is the flattened deformation tensor.
The first variation of $\fF$ is the global residual
\begin{align}
\mathbf R_g=\tdelta \fF=\sum_{i=1}^N V_i \frac{\partial \psi(F_i)}{\partial F_i} \tdelta F_i=\sum_{i=1}^N V_i P_i\tdelta F_i=\sum_{i=1}^N \mathbf R_i\label{eq:R}.
\end{align}
The variation of $\mathbf R_g$ is the global stiffness tangent matrix
\begin{align}
\mathbf K_g=\tdelta\mathbf  R_g=\tdelta^2 \fF=\sum_{i=1}^N V_i \tdelta F_i^T \frac{\partial^2 \psi(F_i)}{\partial F_i^T \partial F_i} \tdelta F_i=\sum_{i=1}^N V_i \tdelta F_i^T \mathbf D \tdelta F_i=\sum_{i=1}^N \mathbf K_i\label{eq:K},
\end{align}

where $V_i$ is the initial nodal volume; $\mathbf R_i,\, \mathbf K_i$ are the nodal residual and nodal tangent stiffness matrix, respectively:
\begin{align}
\mathbf R_i=V_i P_i\tdelta F_i, \, \mathbf K_i=V_i \tdelta F_i^T \mathbf D\tdelta F_i \label{eq:RaKa}.
\end{align} 
The summation of all particles is the global assembling, which is the same as the finite element method. The remaining work is on how to assemble the nodal residual and nodal stiffness matrix. Eq.\ref{eq:RaKa} shows that nodal residual and nodal stiffness are some matrix operations on $\tdelta F$. In the framework of SPH, we have
\begin{align}
\bF_{i}=\sum_{\cS_i}(\by_j-\by_i)\otimes \tnabla W_i(\bx_{ij}) V_j.
\end{align}
The variation of $\tdelta \bF_{i}$ reads
\begin{align}
\tdelta \bF_{i}=\sum_{ \cS_i}(\tdelta \by_j-\tdelta \by_i)\otimes \tnabla W_i(\bx_{ij}) V_j.\label{eq:vaF}
\end{align}
where $ V_j$ is the volume for particle $\bx_j$. For the purpose of numerical implementation, $\tdelta \bF_i$ in 3D can be written as a matrix $\tdelta F_i$ with the dimensions of $9 \times 3 n_{\bx_i}$, which can be assembled with the following order, where $n_{\bx_i}$ is the number of particles in $\cS_i$ (${\bx_i}$ is also included). The assembling process on nodal level is called as nodal assembly.

Assume particle $\bx_i$'s neighbors $N_{\bx_i}=\{j_0,j_1,...,j_k,...,j_{n_i-1}\}$, the first particle $j_0$ denotes the particle $\bx_i$. Here the convention for index starts from 0. For each particle in the neighbor list, we use $R=\tnabla W_i(\bx_{ij}) V_j$, the terms in $R$ can be added to the $\tdelta F$ as 
\begin{align}
\tdelta F_{0,3 k}&=R_0, &\tdelta F_{0,0}=\tdelta F_{0,0}-R_0\notag\\
\tdelta F_{3,3 k}&=R_1, &\tdelta F_{3,0}=\tdelta F_{3,0}-R_1\notag\\
\tdelta F_{6,3 k}&=R_2,&\tdelta F_{6,0}=\tdelta F_{6,0}-R_2\notag\\
\tdelta F_{1,3 k+1}&=R_0, &\tdelta F_{1,1}=\tdelta F_{1,1}-R_0\notag\\
\tdelta F_{4,3 k+1}&=R_1, &\tdelta F_{4,1}=\tdelta F_{4,1}-R_1\notag\\
\tdelta F_{7,3 k+1}&=R_2,&\tdelta F_{7,1}=\tdelta F_{7,1}-R_2\notag\\
\tdelta F_{2,3 k+2}&=R_0, &\tdelta F_{2,2}=\tdelta F_{2,2}-R_0\notag\\
\tdelta F_{5,3 k+2}&=R_1, &\tdelta F_{5,2}=\tdelta F_{5,2}-R_1\notag\\
\tdelta F_{8,3 k+2}&=R_2,&\tdelta F_{8,2}=\tdelta F_{8,2}-R_2\notag,
\end{align}
where $k$ is the index of particle $\bx_j$ in $N_{\bx_i}$. It should be noted that the above derivation is independent with the actually material constitutions, which can be served as a general framework for the implicit analysis using SPH for many materials. 

With $\tdelta F$ and $\mathbf D$ available for any particle $i$, the tangent stiffness matrix $\mathbf K_i$ at a point $i$ in Eq.\ref{eq:RaKa} is $V_i \tdelta F_{i}^T \,\mathbf D_i \,\tdelta F_{i}$, where $\tdelta F_i $ is a matrix of ${9 \times 3 n_{i}}$. The variation of the deformation gradient enables the construction of tangent stiffness matrix being simply the multiplication of some matrices.

The {residual of hourglass energy functional} in Eq.\ref{eq:Phg} and hourglass tangent stiffness matrix in Eq.\ref{eq:Khg} can be obtained with similar procedure. The Dirichlet and Neumann boundary conditions can be applied on the particles, the same as finite element method. After assembling the global stiffness matrix and residual, the solution is obtainable when solving the linear algebra system
\begin{align}
(\mathbf K_g+\mathbf K^{hg})\bu=\mathbf R_g,\notag
\end{align}
where $\mathbf K_g, \mathbf R_g$ are the global stiffness matrix and global residual vector, respectively.

\section{\hl{Material constitutions}}
In this section, we consider only the elastic materials, including the linear elastic material, two hyperelastic materials. The other elastic materials can be formulated with the similar procedure.

The elastic energy density for linear isotropic material is
\begin{align}
\psi(\boldsymbol{\varepsilon})=\frac{1}{2} \lambda(\operatorname{tr} \boldsymbol{\varepsilon})^{2}+\mu \boldsymbol{\varepsilon} : \boldsymbol{\varepsilon}\label{eq:isolinear}
\end{align}
where $\boldsymbol{\varepsilon}=\frac{1}{2}(\bF^T+\bF)-\mathbf I$, $\lambda,\mu$ are the lam\'e constants for isotropic elastic material.

The material tensor $\mathbf D_4=\frac{\partial^2 \psi(\bF_i)}{\partial \bF_i^T \partial \bF_i}$ can be written as matrix form $\mathbf D$
\begin{align}
\mathbf D=\begin{bmatrix}
\lambda+2\mu & 0 & 0 & 0 & \lambda & 0 & 0 & 0 & \lambda \\
 0 & \mu & 0 & \mu & 0 & 0 & 0 & 0 & 0 \\
 0 & 0 & \mu & 0 & 0 & 0 & \mu & 0 & 0 \\
 0 & \mu & 0 & \mu & 0 & 0 & 0 & 0 & 0 \\
 \lambda & 0 & 0 & 0 & \lambda+2\mu & 0 & 0 & 0 & \lambda \\
 0 & 0 & 0 & 0 & 0 & \mu & 0 & \mu & 0 \\
 0 & 0 & \mu & 0 & 0 & 0 & \mu & 0 & 0 \\
 0 & 0 & 0 & 0 & 0 & \mu & 0 & \mu & 0 \\
 \lambda & 0 & 0 & 0 & \lambda & 0 & 0 & 0 & \lambda+2\mu 
\end{bmatrix}.
\end{align}

For the case of Neo-Hooke material \cite{simo1985variational}, the strain energy can be expressed as
\begin{align}
\psi(\bF)=\frac{1}{2} \kappa (J-1)^2+\frac{1}{2} \mu ({J^{-2/3}}\bF:\bF-3).\label{eq:nhm}
\end{align}
The first Piola-Kirchoff stress is
\begin{align}
\mathbf P=\frac{\partial \psi(\bF)}{\partial \bF}=\frac{\mu}{J^{2/3}} \bF+\Big((J-1)\kappa-\mu \frac{\bF:\bF}{3 J^{5/3}} \Big)J_{,\bF}.
\end{align}
The material tensor can be written as
\begin{align}
&\mathbf D=\fracp{P}{F}=\frac{\mu}{J^{2/3}}\mathbf I_{9\times 9}-\frac{2 \mu}{3 J^{5/3}}(F\otimes J_{,F}+J_{,F}\otimes F)+\notag\\&\big(\kappa+\frac{5\mu}{9 J^{8/3}}\bF:\bF\big)J_{,F}\otimes J_{,F}+\Big(\kappa(J-1)-\frac{\mu}{3 J^{5/3}}\bF:\bF\Big)J_{,FF}
\end{align}
where $J_{,F}$ is the vector form of $J_{,\mathbf F}$, and 
\begin{align}J_{,F F}=
\left[
\begin{array}{ccccccccc}
 0 & 0 & 0 & 0 & F_{9} & \text{-}F_{8} & 0 & \text{-}F_{6} & F_{5} \\
 0 & 0 & 0 & \text{-}F_{9} & 0 & F_{7} & F_{6} & 0 & \text{-}F_{4} \\
 0 & 0 & 0 & F_{8} & \text{-}F_{7} & 0 & \text{-}F_{5} & F_{4} & 0 \\
 0 & \text{-}F_{9} & F_{8} & 0 & 0 & 0 & 0 & F_{3} & \text{-}F_{2} \\
 F_{9} & 0 & \text{-}F_{7} & 0 & 0 & 0 & \text{-}F_{3} & 0 & F_{1} \\
 \text{-}F_{8} & F_{7} & 0 & 0 & 0 & 0 & F_{2} & \text{-}F_{1} & 0 \\
 0 & F_{6} & \text{-}F_{5} & 0 & \text{-}F_{3} & F_{2} & 0 & 0 & 0 \\
 \text{-}F_{6} & 0 & F_{4} & F_{3} & 0 & \text{-}F_{1} & 0 & 0 & 0 \\
 F_{5} & \text{-}F_{4} & 0 & \text{-}F_{2} & F_{1} & 0 & 0 & 0 & 0 \\
\end{array}
\right].
\end{align}

Another energy density functional for the compressible neo-Hookean material \cite{bonet1997nonlinear} is
\begin{align}
\psi=\frac{\mu}{2}\left(\bF:\bF-3\right)-\mu \ln J+\frac{\lambda}{2}(\ln J)^{2} \label{eq:neohook2}
\end{align}
The first Piola-Kirchhoff stress and the material matrix are
\begin{align}
\mathbf P&=\mu \mathbf F-\mu\frac{J_{,\bF}}{J}+\lambda \frac{\ln J}{J}J_{,\bF}\\
\mathbf D&=\mu \mathbf I_{9\times 9}+\lambda \frac{\ln J}{J} J_{,FF}+\frac{1}{J^2}(\lambda-\lambda\ln J+\mu)J_{,F}\otimes J_{,F}
\end{align}
Eq.\ref{eq:neohook2} is used to model the rubber in section \ref{sec:rubberPull}.

\section{Numerical examples}\label{sec:NumericalTest}
We give \hl{six} numerical examples to validate the implicit formulation of dual-support SPH and test the performance of the hourglass control. The numerical results are compared with the theoretical solutions or that by finite element method. \hl{Traditionally, SPH is solved by explicit integration methods, such as Velocity-Verlet algorithm, Leapfrog integration. Explicit integration method is limited by the maximal time increment for the reason of numerical stability. The maximal time increment depends on the minimal particle size $\Delta x_{min}$ and the wave sound speed ($C$)in the media, i.e. $\Delta t_{max}\leq\Delta x_{min}/C$. The computer cost is economical for short-duration models but is very expensive for long-term models. Implicit algorithm is unconditioned stable for any time increment, thus has advantage for solving static or quasi-static problems. Traditional SPH can't be solved implicitly because the tangent stiffness matrix is not available. The variational formulation of SPH in this paper obtains the residual and tangent stiffness matrix with ease, and thus provides great feasibility for implicit analysis by SPH for static/dynamic problems. For dynamic problem,  Hilber-Hughes-Taylor (HHT) integration \cite{Hilber1977Jul}, Newmark method \cite{newmark1959method} can be readily used. For simplify, we assume the acceleration in Eq.\ref{eq:gsph} to be zero and study a series of static problems.}
%
%
\subsection{3D Cantilever loaded at the end}\label{sbsec:3Dbeam}
\begin{figure}
	\centering
		\includegraphics[width=8cm]{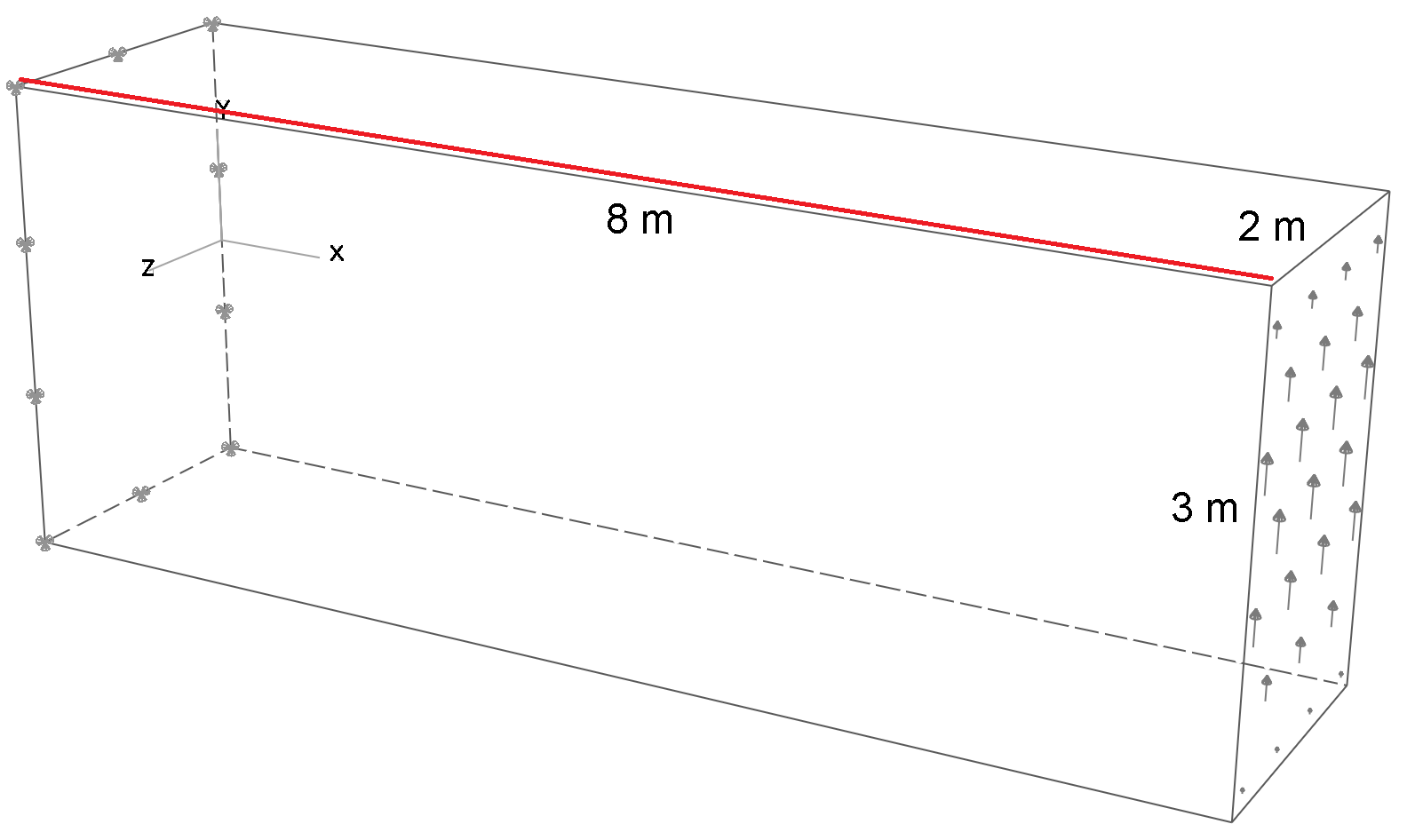}
	\caption{Setup of the thick beam}
	\label{fig:thickBeam}
\end{figure}

A three-dimensional cantilever beam loaded at the end with pure shear traction force is considered. The beam with dimensions of height of $D=3$m, length of $L=8$ m and thickness of $t=2$ m and shear load of parabola distribution is shown in Fig.\ref{fig:thickBeam}. The analytical solution for the beam is \cite{timoshenko412theory,zhuang2010aspects}
\begin{align}
u_x&=\frac{P y}{6 EI}\big[(6 L-3 x)x+(2+\nu)(y^2-\frac{D^2}{4})\big]\label{eq:cantileverux}\\
u_y&=-\frac{P}{6 EI}\big[3 \nu y^2(L-x)+(4+5 \nu)\frac{D^2 x}{4}+(3L-x)x^2 \big]\label{eq:cantileveruy}\\
\sigma_{xx}(x,y)&=\frac{P(L-x)y}{I},\sigma_{yy}(x,y)=0,\tau_{xy}(x,y)=-\frac{P}{2 I}\big(\frac{D^2}{4}-y^2\big),\label{eq:cantileversigma}
\end{align}
where $P=-1000$ N,$I=\frac{D^3}{12}$.
The material parameters are taken as $E=30 $GPa,$\nu=0.3$. The particles on the left boundary are constrained by the exact displacements from Eq.\ref{eq:cantileverux} and Eq.\ref{eq:cantileveruy} and the loading on the right boundary follows Eq.\ref{eq:cantileversigma}.
The error norm in displacement for particle $i$ is calculated by
\begin{align}
\|u\|_{error}=\sqrt{\frac{\sum_{i=1}^{N} (\mathbf u_i-\mathbf u_i^h)\cdot (\mathbf u_i-\mathbf u_i^h)V_i}{\sum_{i=1}^{N} \mathbf u_i\cdot \mathbf u_i V_i}}\label{eq:uerror}
\end{align}
The exact strain energy and numerical strain energy are computed by
\begin{align}
E_{numerical}&=\frac{1}{2}\sum_{i=1}^{N} \varepsilon_i^T \mathbf D\varepsilon_i V_i\notag\\
E_{exact}&=\frac{1}{2}\int_{\Omega} \varepsilon_i^T \mathbf D\varepsilon_i\mathrm{d}V_i\notag\\
E_{error}&=\frac{E_{numerical}}{E_{exact}}-1\label{eq:Eerror},
\end{align}
where $\mathbf D$ is the material tensor.

\begin{figure}
	\centering
		\includegraphics[width=10cm]{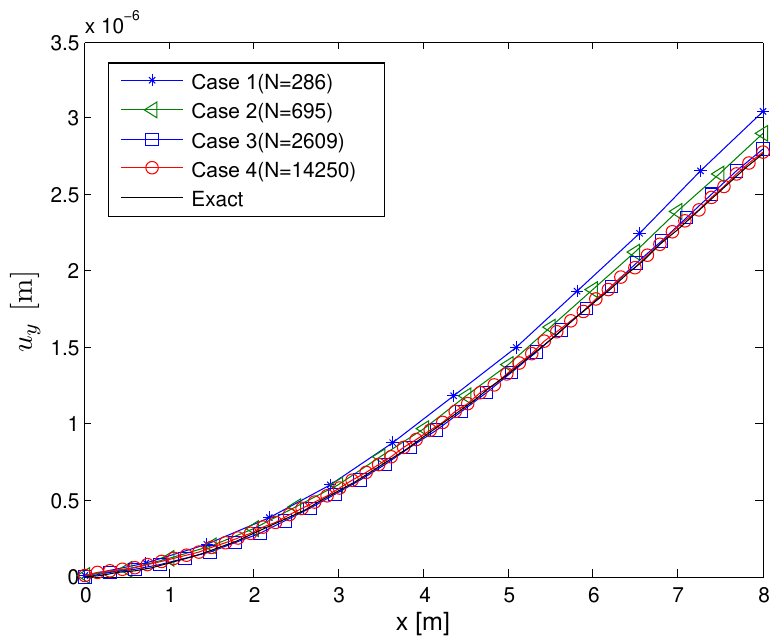}
	\caption{Displacement curve in $y$ direction for different discretizations with hourglass control.}
	\label{fig:thickBeamUy}
\end{figure}

We tested four cases with different discretizations. The statistics of the particle number, the supports and the displacement error and energy are given in Table.\ref{tab:beamStatistic}. It can be seen that the numerical results converge to the theoretical solution with the increase of the number of particles. The $y$-displacements of particles on the red line in Fig.\ref{fig:thickBeam} are plotted with good agreement to theoretical solution in Fig.\ref{fig:thickBeamUy}. 

\begin{table}[h]
\begin{center}
\begin{tabular}{ | c|c | c | c | c | c|c|c|c|c| }
\hline
Case & $N$ &$h_{min}$ & $h_{max}$ & $\|u\|_{error}$ & $E_{strain}$ & $\fF^{hg}$ & $e(E_{strain})$\\ \hline
1 & 286&0.844&2.235&0.0859&3.011e-3&2.422e-6 & 0.0859
\\ \hline
2 & 695&0.608&1.691&0.0729&2.948e-3&1.487e-6&0.0631
\\ \hline
3 & 2609&0.316&1.102&0.0273&2.891e-3&7.785e-7&0.0426
\\ \hline
4 & 14250 & 0.164& 0.675&0.0208&2.819e-3&3.625e-7&0.0165\\ \hline
\end{tabular}
\caption{Convergence study for different discretizations, where $e(E)=|E-E^{exact}|/|E^{exact}|$, $N$ is the number of particles. The exact strain energy is $E_{strain}^{exact}=0.00277284$.}\label{tab:beamStatistic}
\end{center}
\end{table}


%
\subsection{Plate under compression}
\begin{figure}
	\centering
		\includegraphics[width=.5\textwidth]{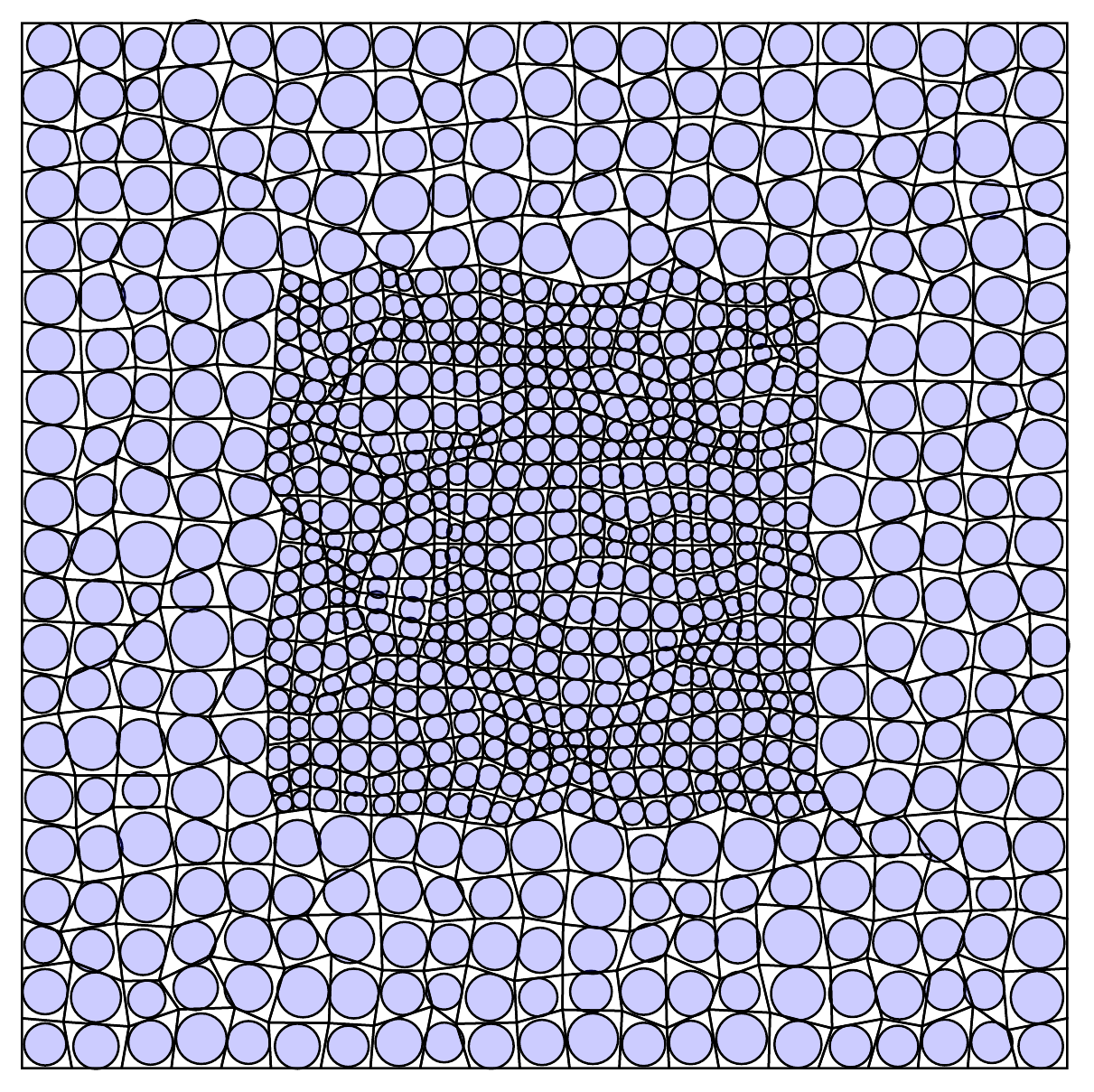}
	\caption{Distribution of the particles, the radius represents the size of the particle.}
	\label{fig:pdistri}
\end{figure}

\begin{figure}
 \centering
 \subfigure[$u_x$ when $\alpha=0$]{
 \label{fig:xhg0}
 \includegraphics[width=.4\textwidth]{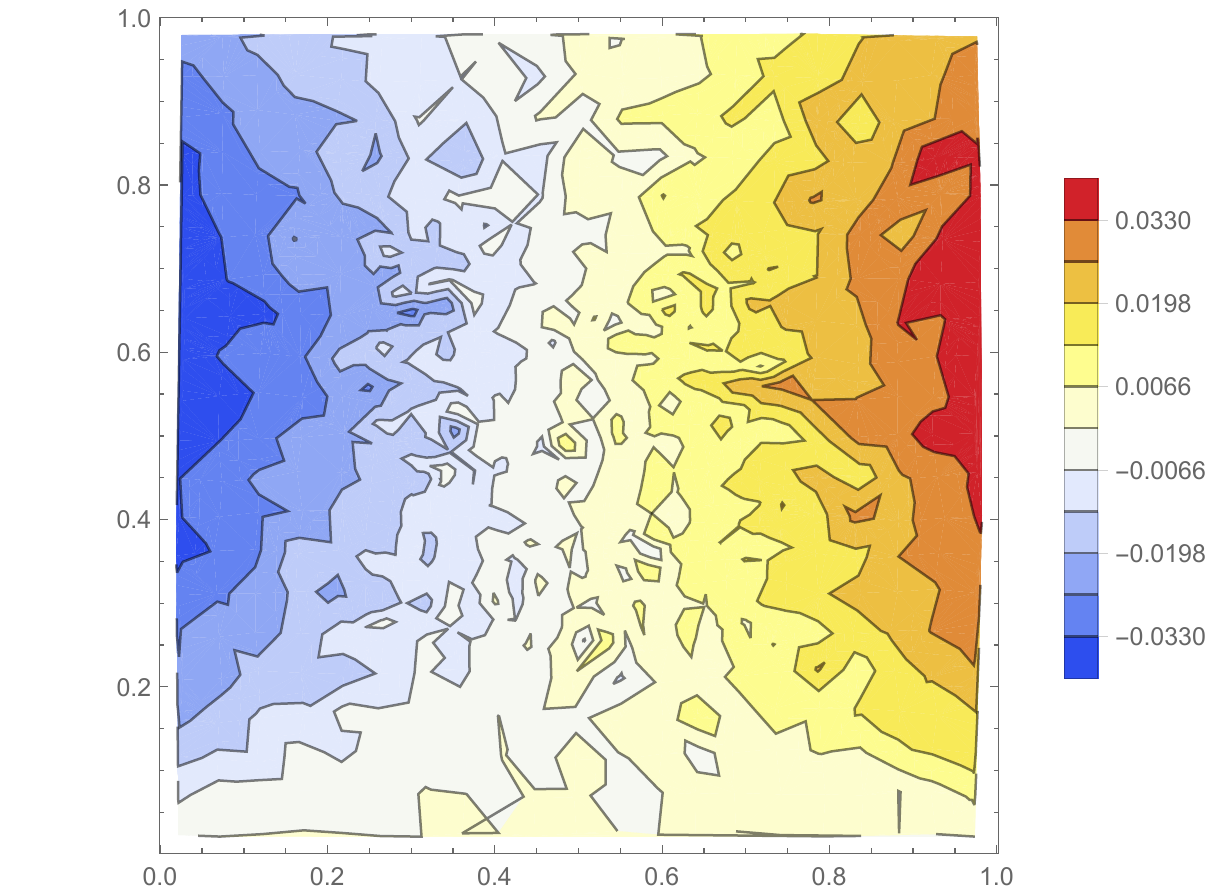}}
 \vspace{.1in}
 \subfigure[$u_y$ when $\alpha=0$]{
 \label{fig:yhg0}
 \includegraphics[width=.4\textwidth]{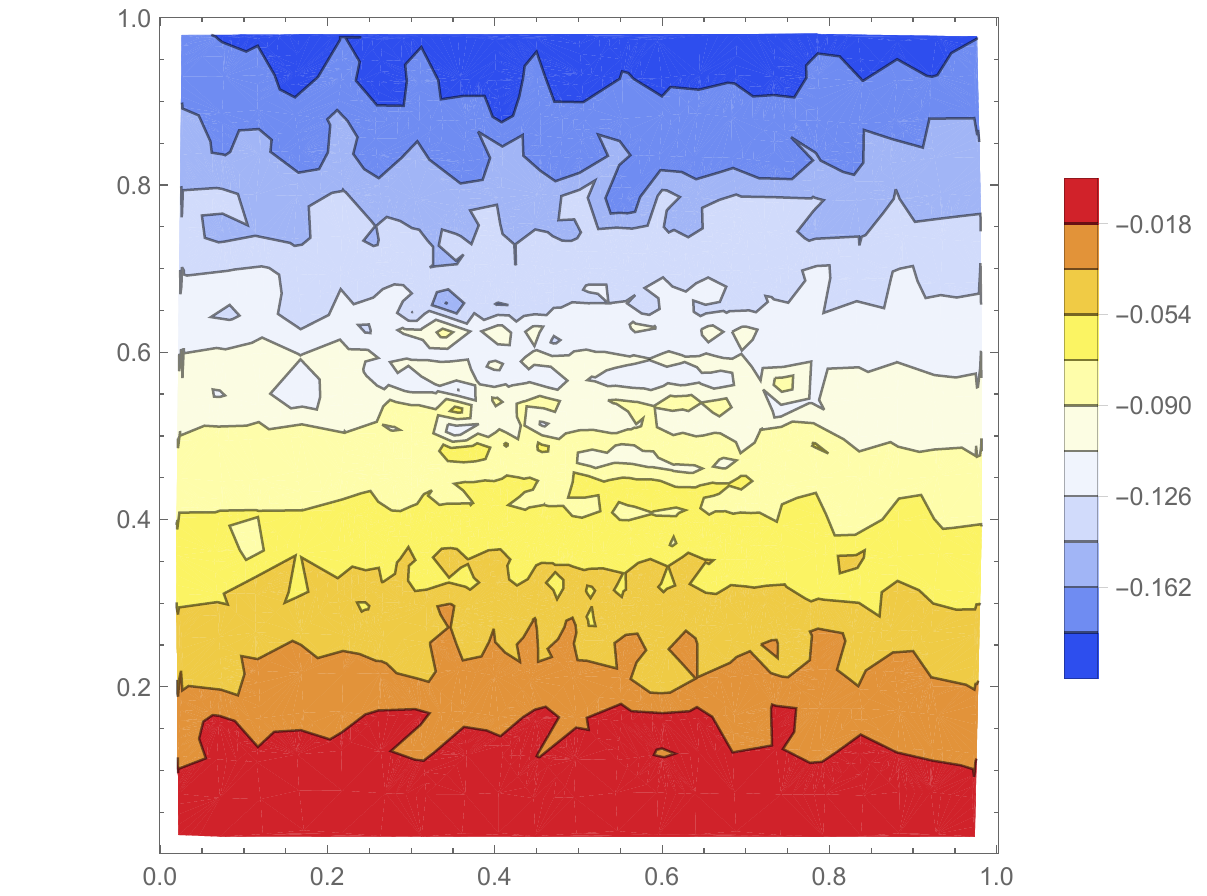}}\\
 \vspace{.1in}
 \subfigure[$u_x$ when $\alpha=0.2$]{
 \label{fig:xhgd2}
 \includegraphics[width=.4\textwidth]{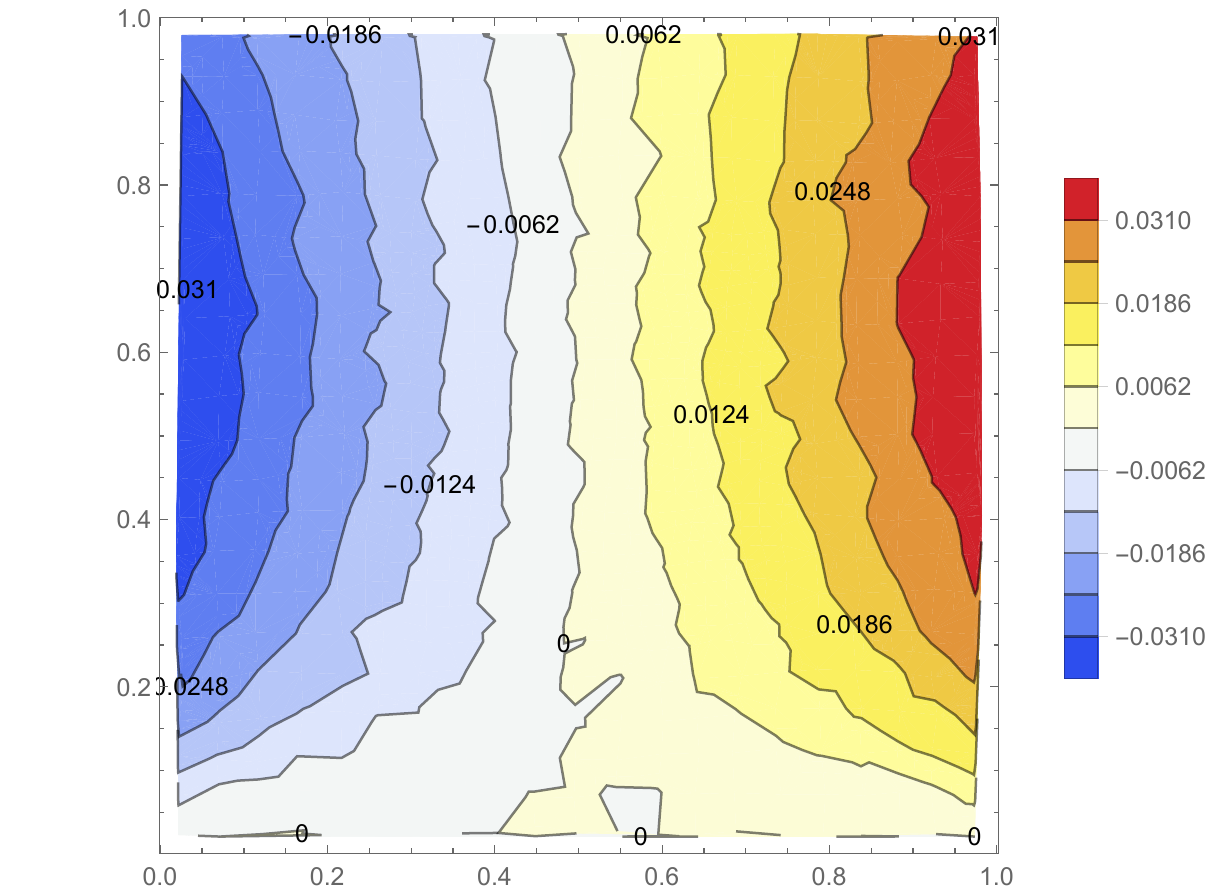}}
 \vspace{.1in}
 \subfigure[$u_y$ when $\alpha=0.2$]{
 \label{fig:yhgd2}
 \includegraphics[width=.4\textwidth]{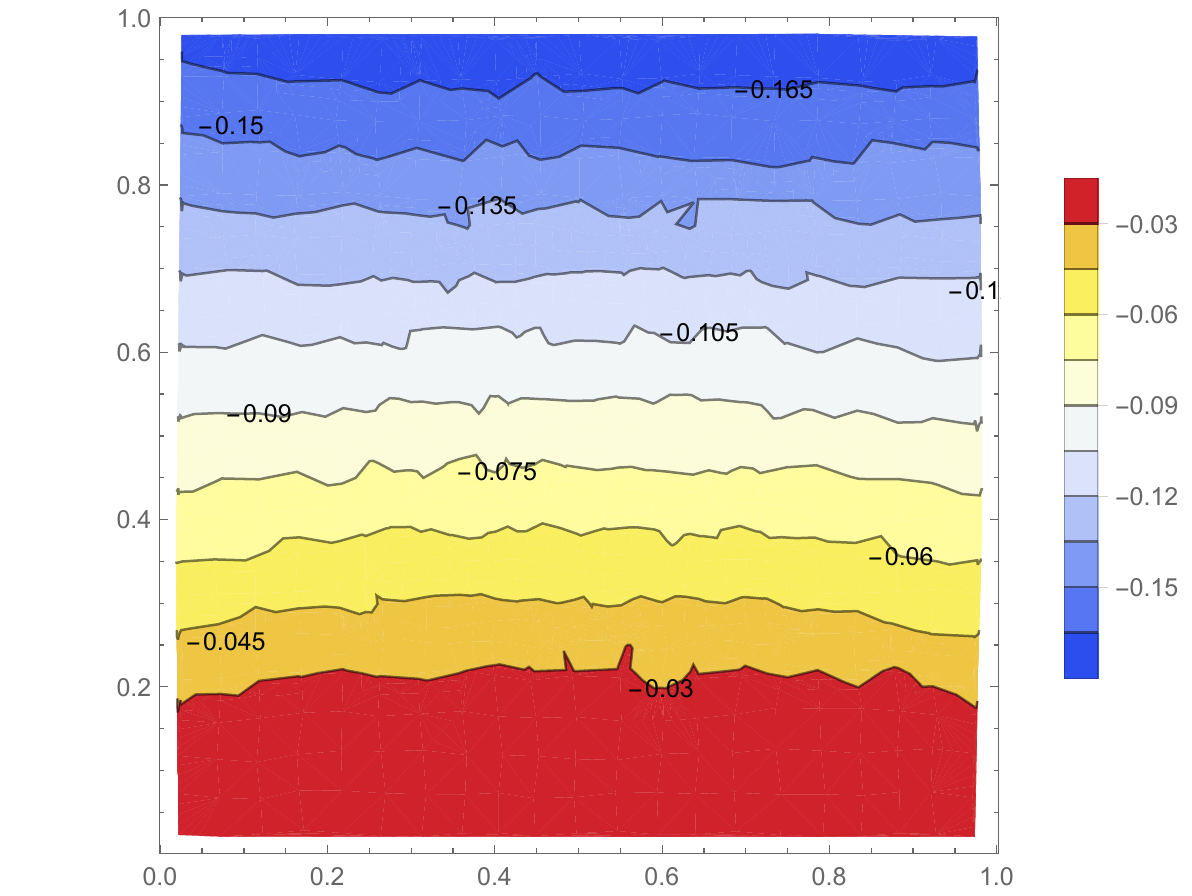}}\\
 \vspace{.1in}
 \subfigure[$u_x$ when $\alpha=2$]{
 \label{fig:xhg2}
 \includegraphics[width=.4\textwidth]{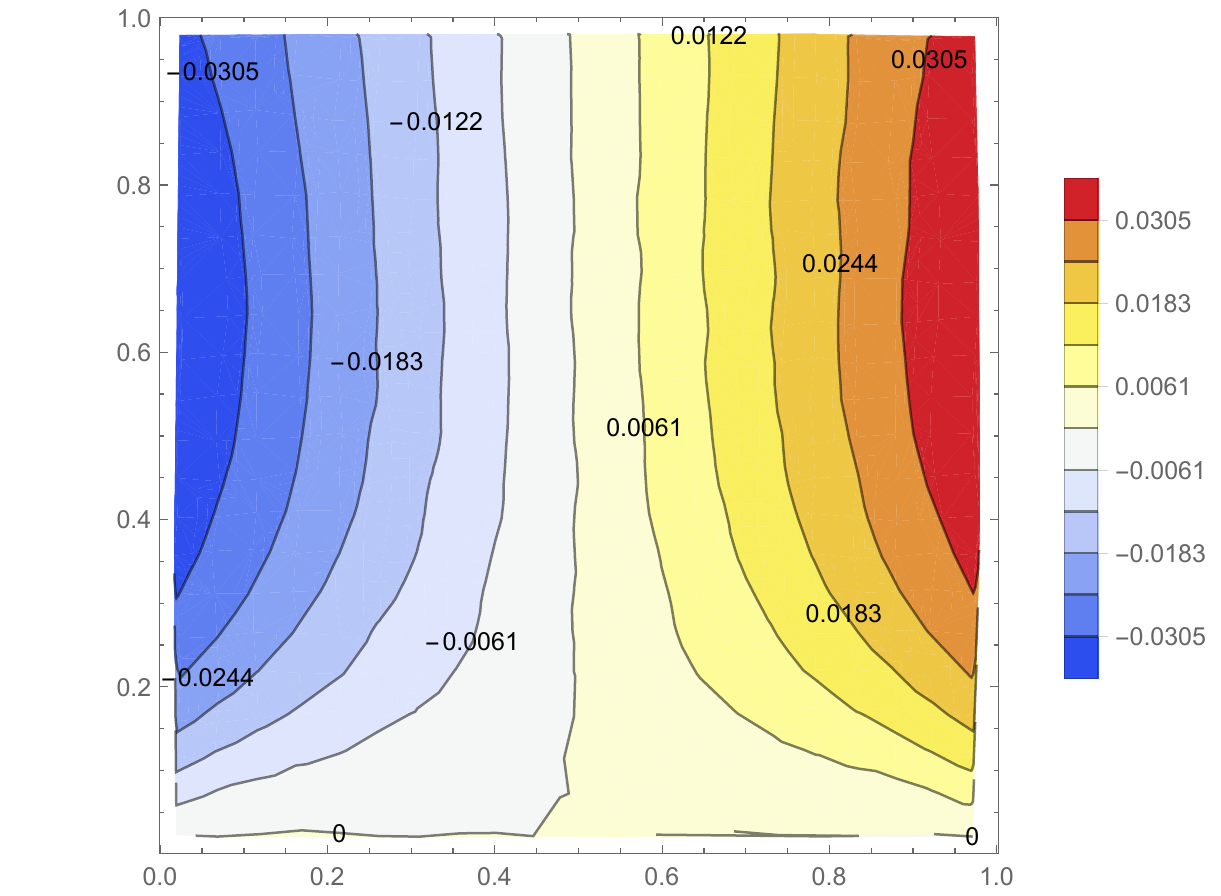}}
 \vspace{.1in}
 \subfigure[$u_y$ when $\alpha=2$]{
 \label{fig:yhg2}
 \includegraphics[width=.4\textwidth]{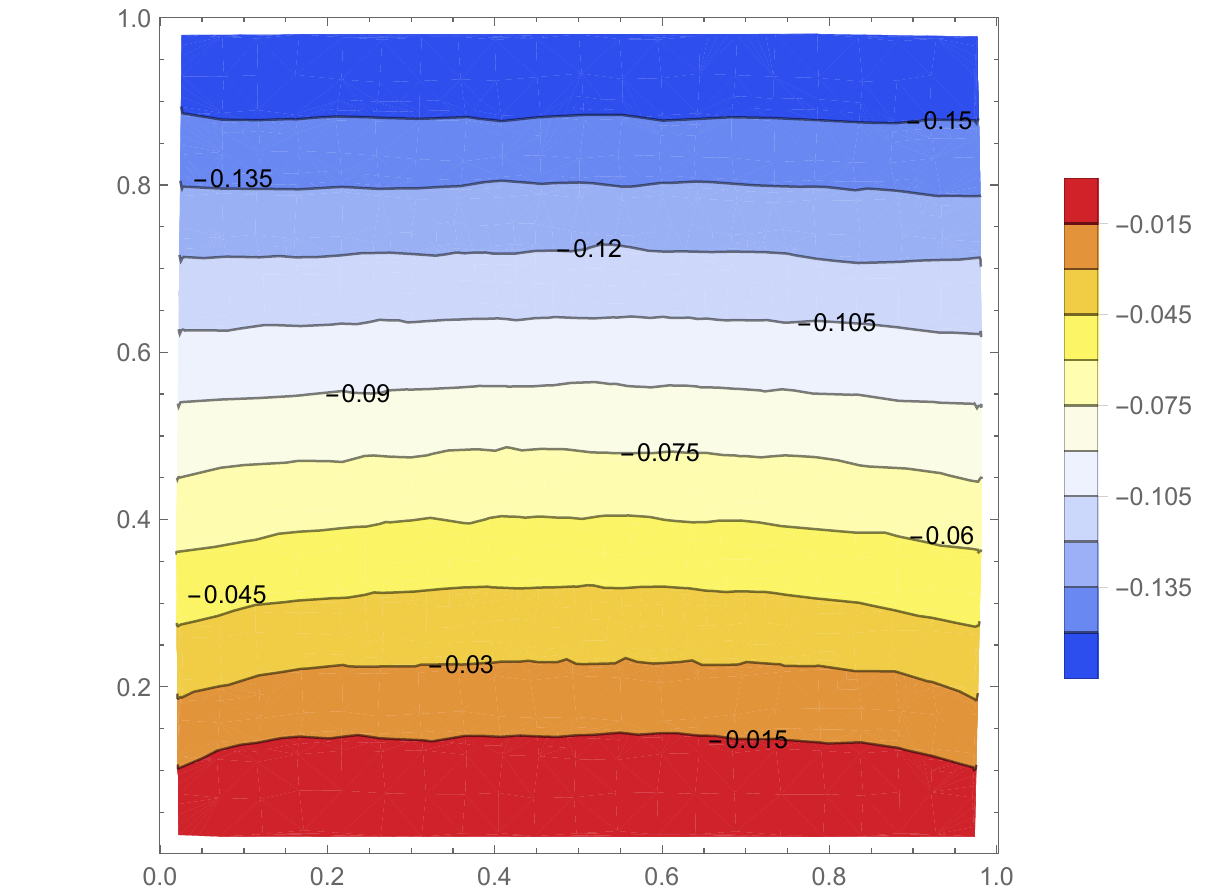}}\\
 \vspace{.1in}
\caption{Contour of displacement field for different hourglass penalties.}
\label{fig:hgUx100}
\end{figure}

\begin{figure}
 \centering
 \subfigure[$u_x$ when $\alpha=10$]{
 \label{fig:xhg10}
 \includegraphics[width=.4\textwidth]{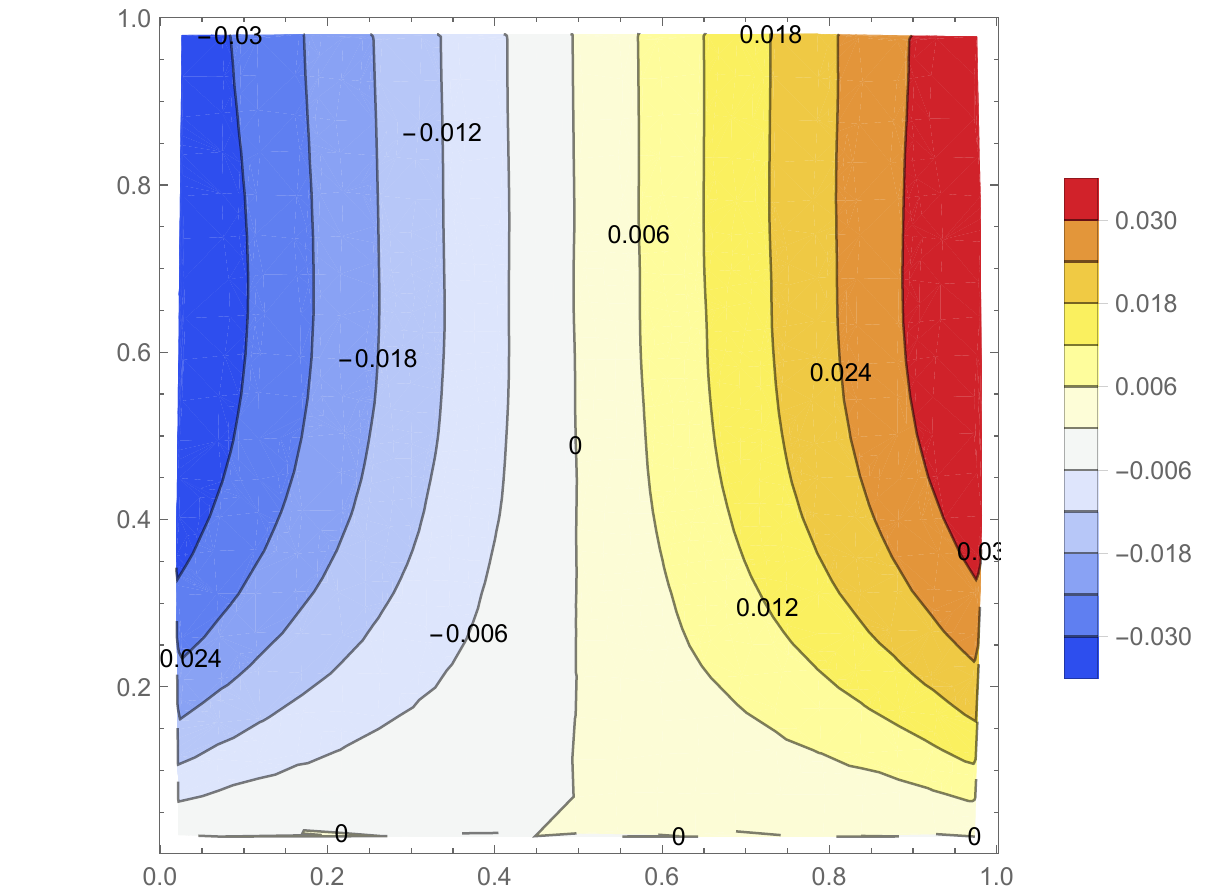}}
 \vspace{.1in}
 \subfigure[$u_y$ when $\alpha=10$]{
 \label{fig:yhg10}
 \includegraphics[width=.4\textwidth]{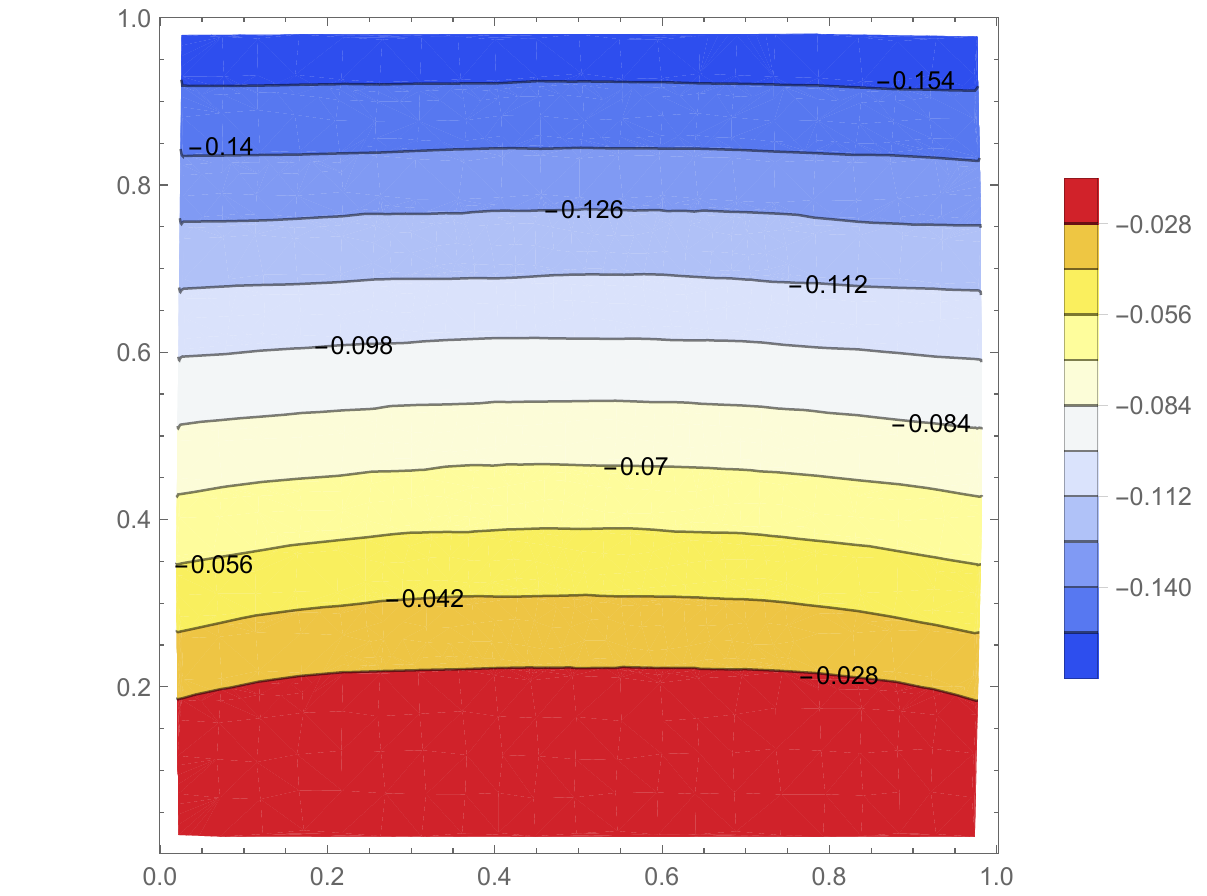}}\\
 \vspace{.1in}
 \subfigure[$u_x$ when $\alpha=100$]{
 \label{fig:xhg100}
 \includegraphics[width=.4\textwidth]{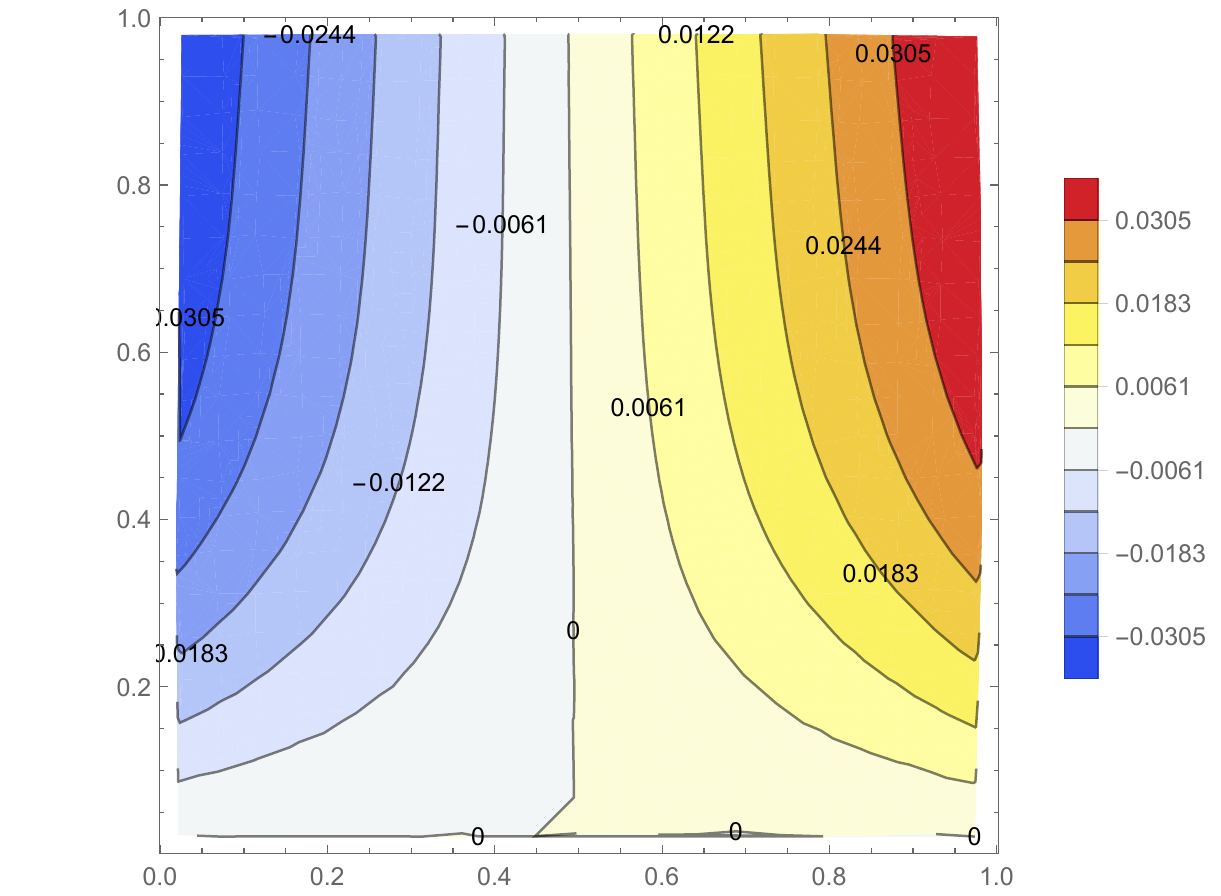}}
 \vspace{.1in}
 \subfigure[$u_y$ when $\alpha=100$]{
 \label{fig:yhg100}
 \includegraphics[width=.4\textwidth]{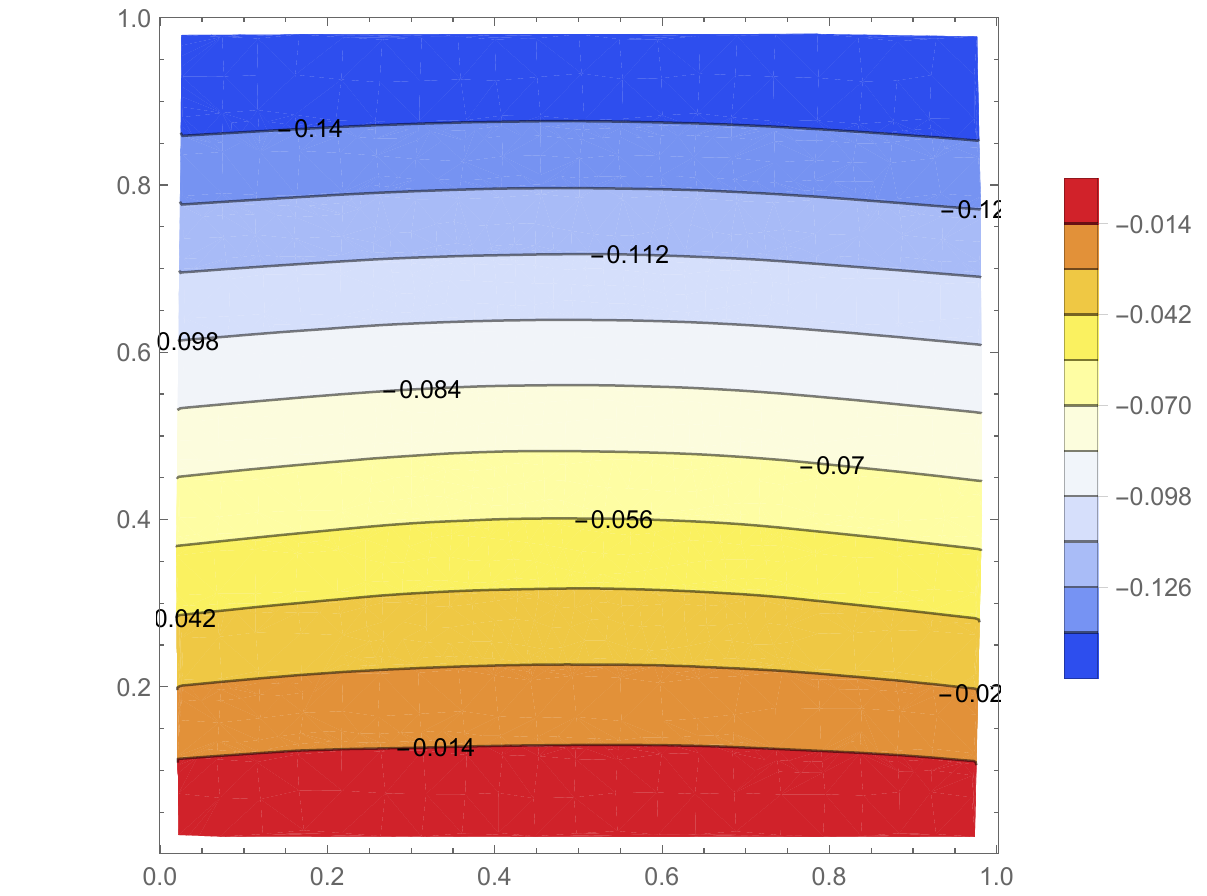}}\\
 \vspace{.1in}
 \subfigure[$u_x$ when $\alpha=10^5$]{
 \label{fig:xhg1e5}
 \includegraphics[width=.4\textwidth]{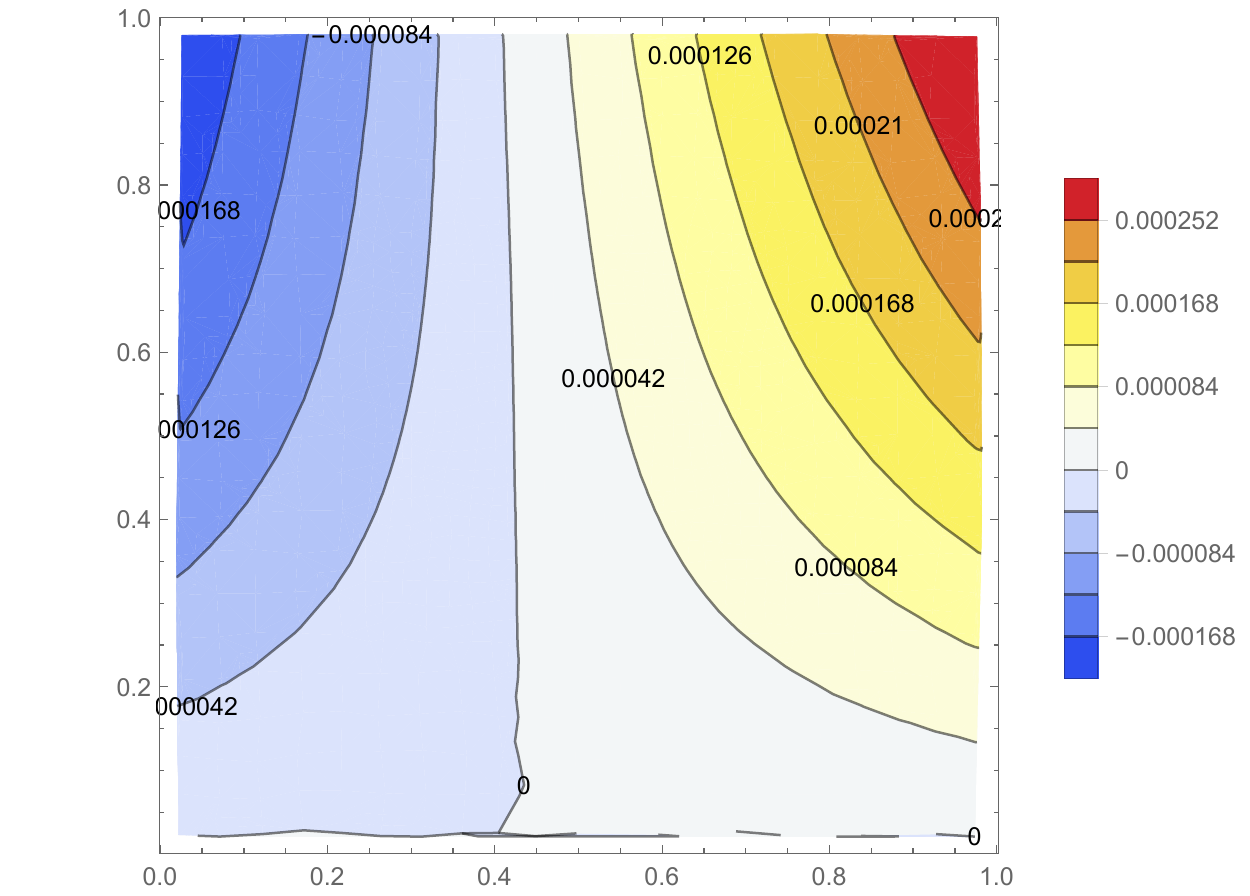}}
 \vspace{.1in}
 \subfigure[$u_y$ when $\alpha=10^5$]{
 \label{fig:yhg1e5}
 \includegraphics[width=.4\textwidth]{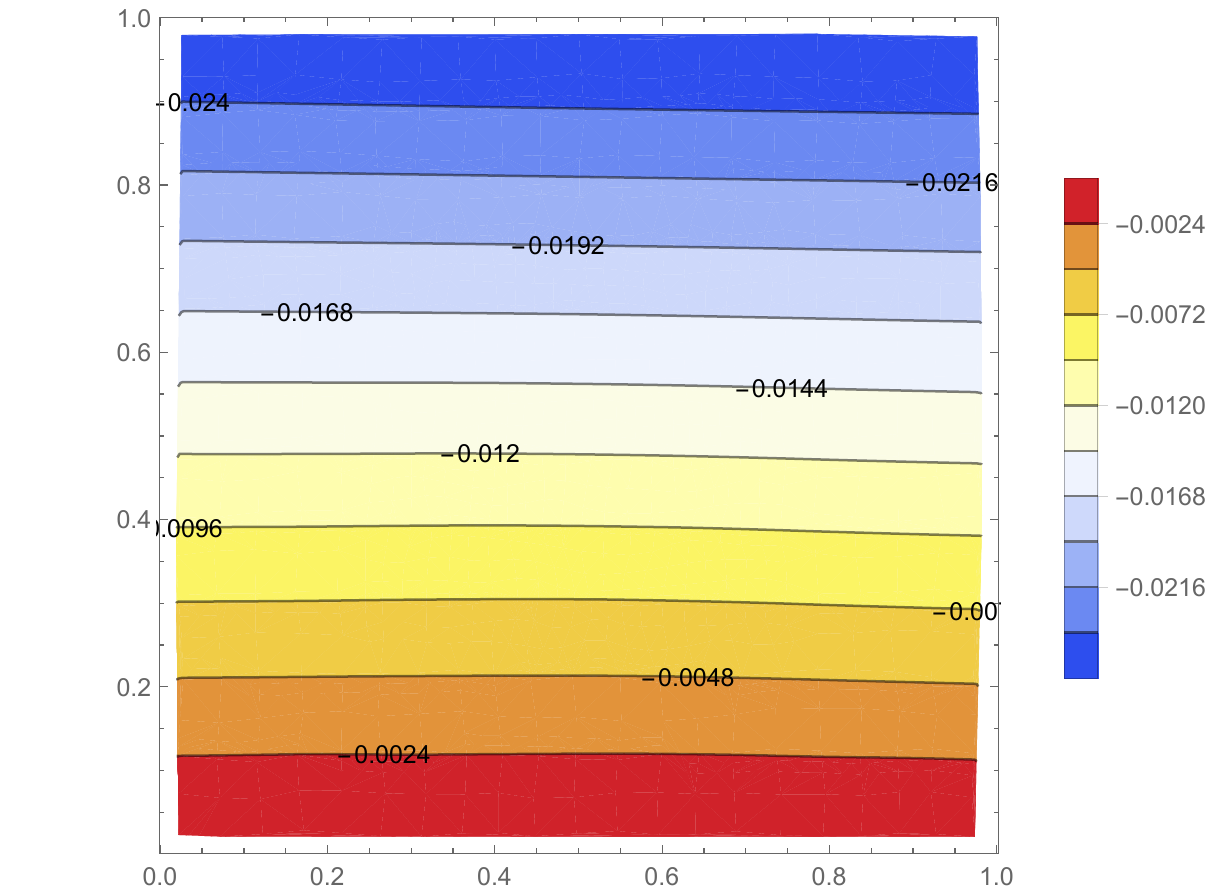}}\\
 \vspace{.1in}
\caption{Contour of displacement field for different hourglass penalties.}
\label{fig:hgUx200}
\end{figure}

\hl{In order to study the influence of variable smoothing length and hourglass penalty, we model a $1\times 1$ m$^2$ plate with material parameters elastic modulus $E=2 $Pa, Poisson ratio $\nu=0.3$. Plane strain condition and linear elasticity in Eq.\ref{eq:isolinear} are assumed. The particles in the bottom are fixed in all directions and the top boundary of the plate is applied with pressure $p=2$ N/m. The plate is discretized with irregular quadrilateral element and then the element is converted into particle positioned at the barycenter of the element. The area of the particle is determined by that of the element. The distribution of the particles is shown in Fig.\ref{fig:pdistri}. The support domain for each particle comprises 10 nearest particles and the smoothing length is selected as the maximal distance of neighbors with respect to the particle. We test different hourglass penalty $\alpha\in\{0,0.2, 2, 10, 100, 10^3, 10^5\}$. The contour plots of the displacement field are shown in Figs.(\ref{fig:hgUx100}, \ref{fig:hgUx200}). The maximal displacement in $x,y-$directions are given in Table .\ref{tab:uxyMaxp}. It can be seen that the hourglass control has very positive effect on the formulation. Without hourglass control, the displacement field is quite poor. On the other hand, too large hourglass penalty would make the mechanical system over-stiff, as shown in Figs.(\ref{fig:xhg1e5},\ref{fig:yhg1e5}).}
\begin{table}[h]
\begin{center}
\begin{tabular}{ | c|c | c | c | c | c|c|c|c| }
\hline
& $\alpha=0$ & $\alpha=0.2$ & $\alpha=2$ & $\alpha=10$ & $\alpha=100$ & $\alpha=10^3$ & $\alpha=10^5$ & Abaqus\\\hline
max($u_x$) & 0.0393 & 0.0382 & 0.0372& 0.037& 0.0381& 0.0288& 0.000303& 0.04\\
max($u_y$) & -0.202 & -0.175 & -0.1664& -0.16367& -0.1604& -0.1465& -0.0267 & -0.178\\\hline
\end{tabular}
\caption{Maximal displacement for different hourglass penalties.}\label{tab:uxyMaxp}
\end{center}
\end{table}

\subsection{3D Cantilever Tension test}
\begin{figure}
 \centering
 \subfigure[Fine mesh without hourglass control]{
 \label{fig:hgUx1}
 \includegraphics[width=.4\textwidth]{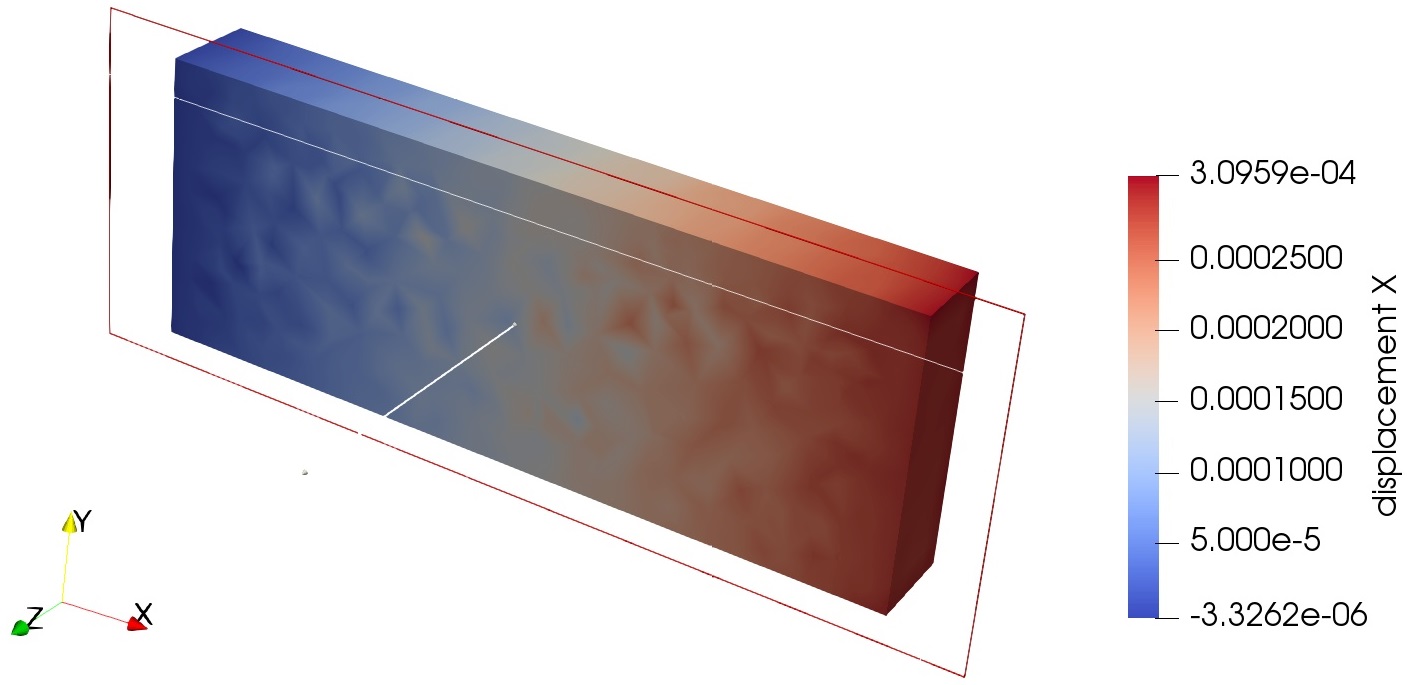}}
 \vspace{.1in}
 \subfigure[Fine mesh with hourglass control]{
 \label{fig:hgUx2}
 \includegraphics[width=.4\textwidth]{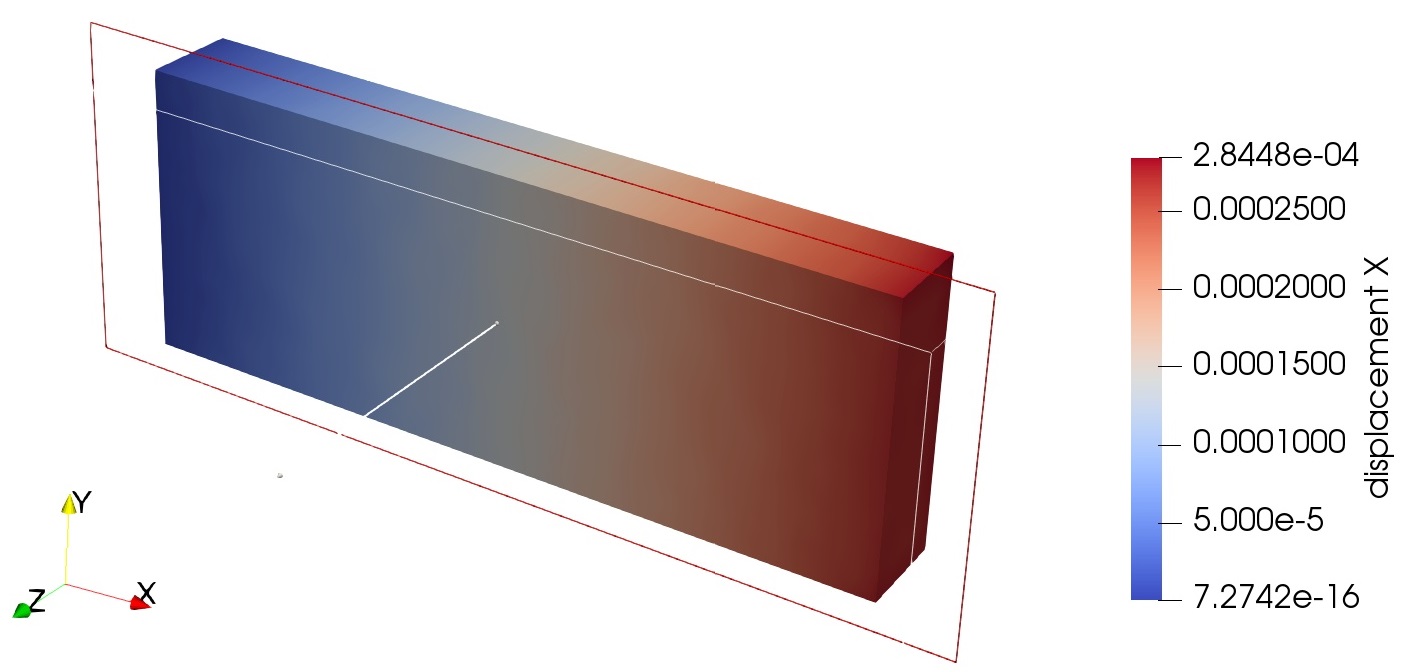}}
 \vspace{.1in}
 \subfigure[Finer mesh without hourglass control]{
 \label{fig:hgUx3}
 \includegraphics[width=.4\textwidth]{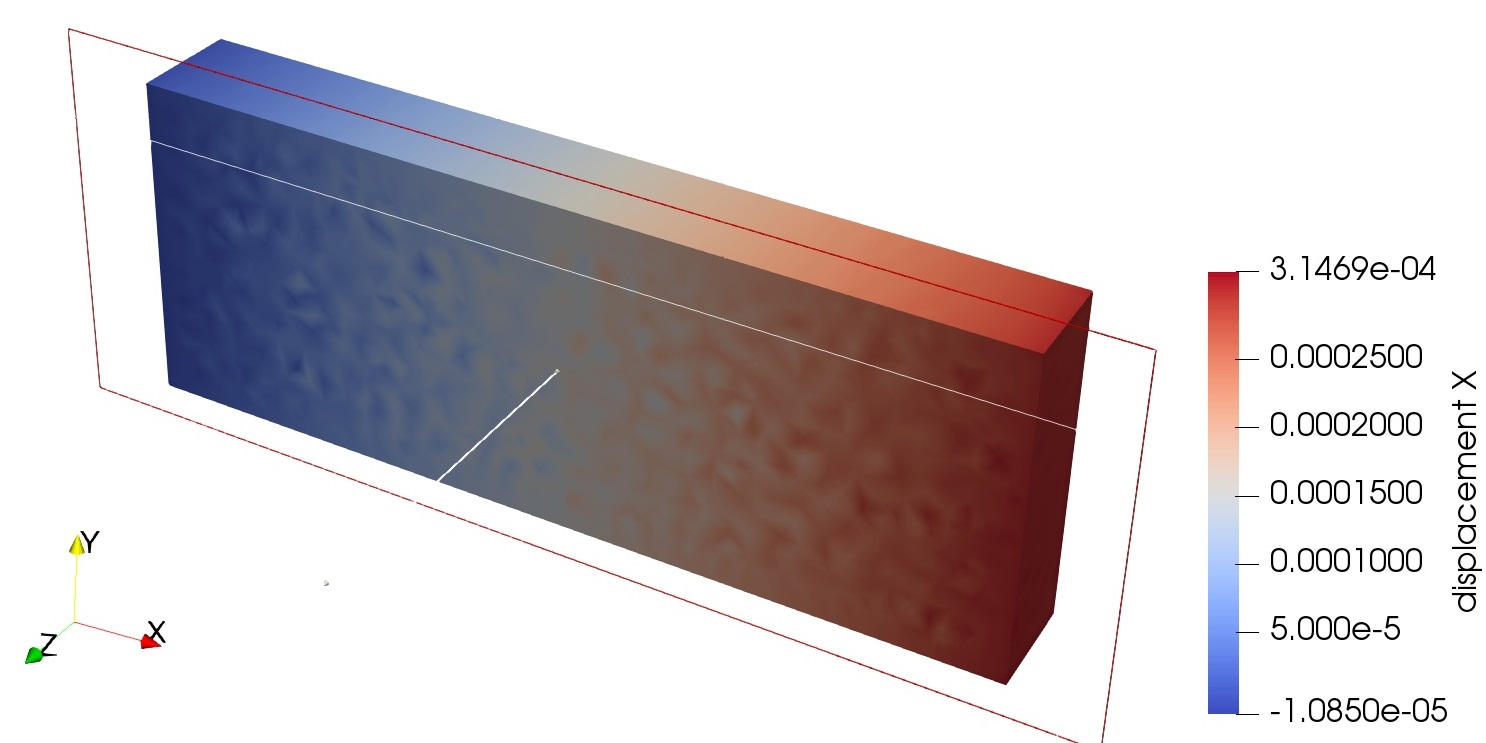}}
 \vspace{.1in}
 \subfigure[Finer mesh with hourglass control]{
 \label{fig:hgUx4}
 \includegraphics[width=.4\textwidth]{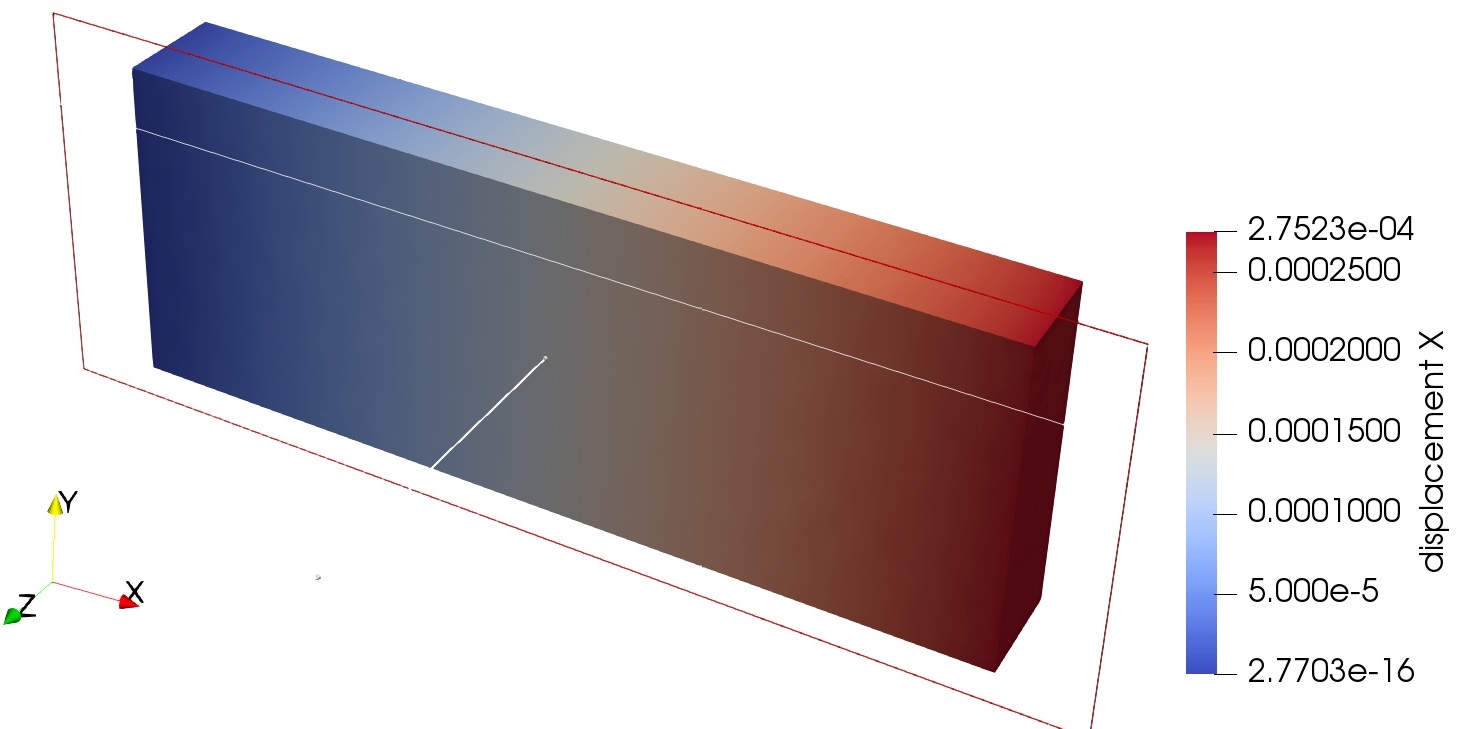}}\\
 \vspace{.1in}
\caption{$x$ displacement on the clip of $z=1$ m.}
\label{fig:hgUx}
\end{figure}

\begin{figure}
 \centering
 \subfigure[Fine mesh without hourglass control]{
 \label{fig:hgDensity1}
 \includegraphics[width=.4\textwidth]{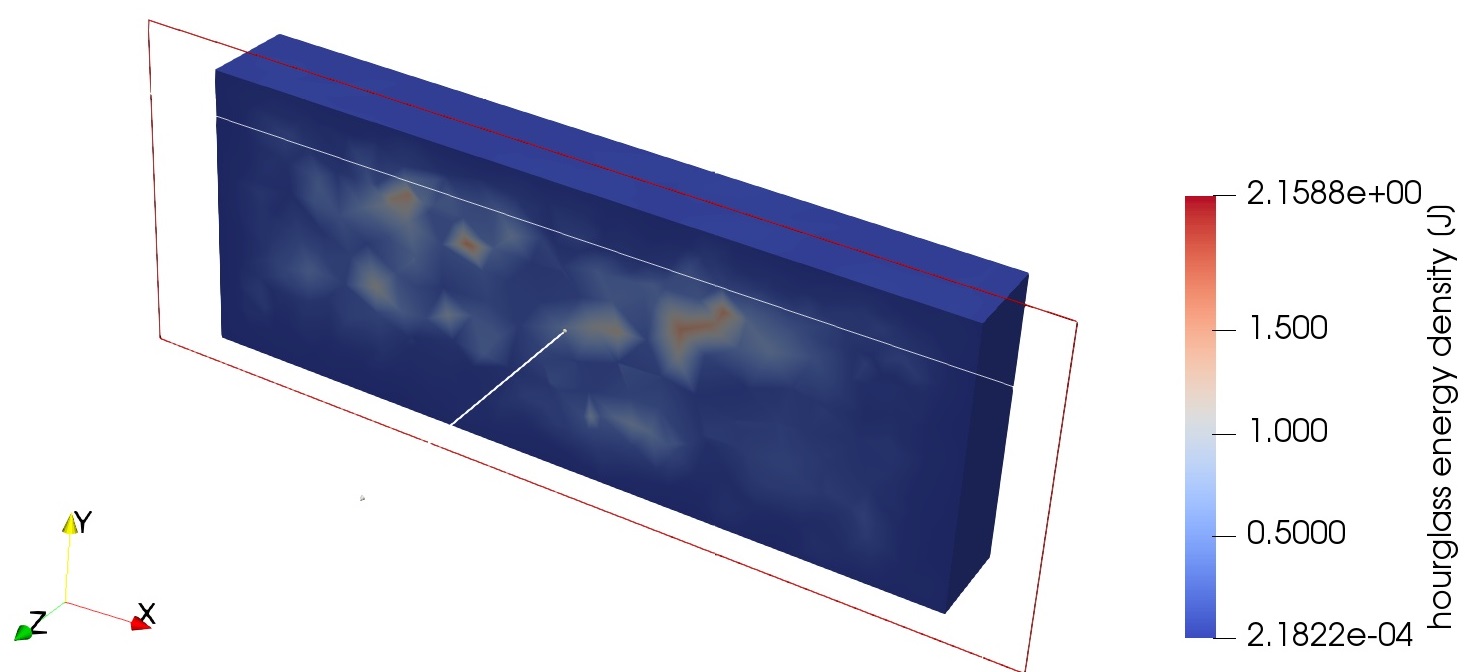}}
 \vspace{.1in}
 \subfigure[Fine mesh with hourglass control]{
 \label{fig:hgDensity2}
 \includegraphics[width=.4\textwidth]{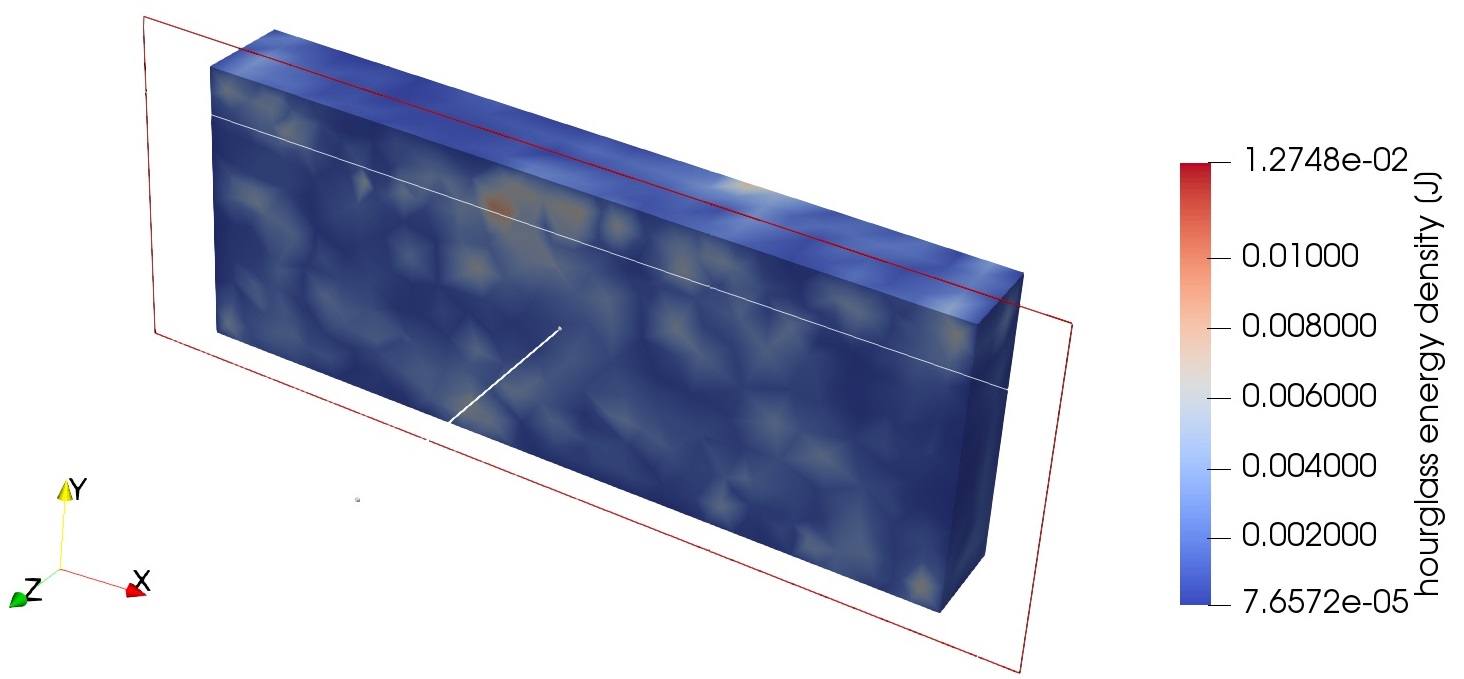}}
 \vspace{.1in}
 \subfigure[Finer mesh without hourglass control]{
 \label{fig:hgDensity3}
 \includegraphics[width=.4\textwidth]{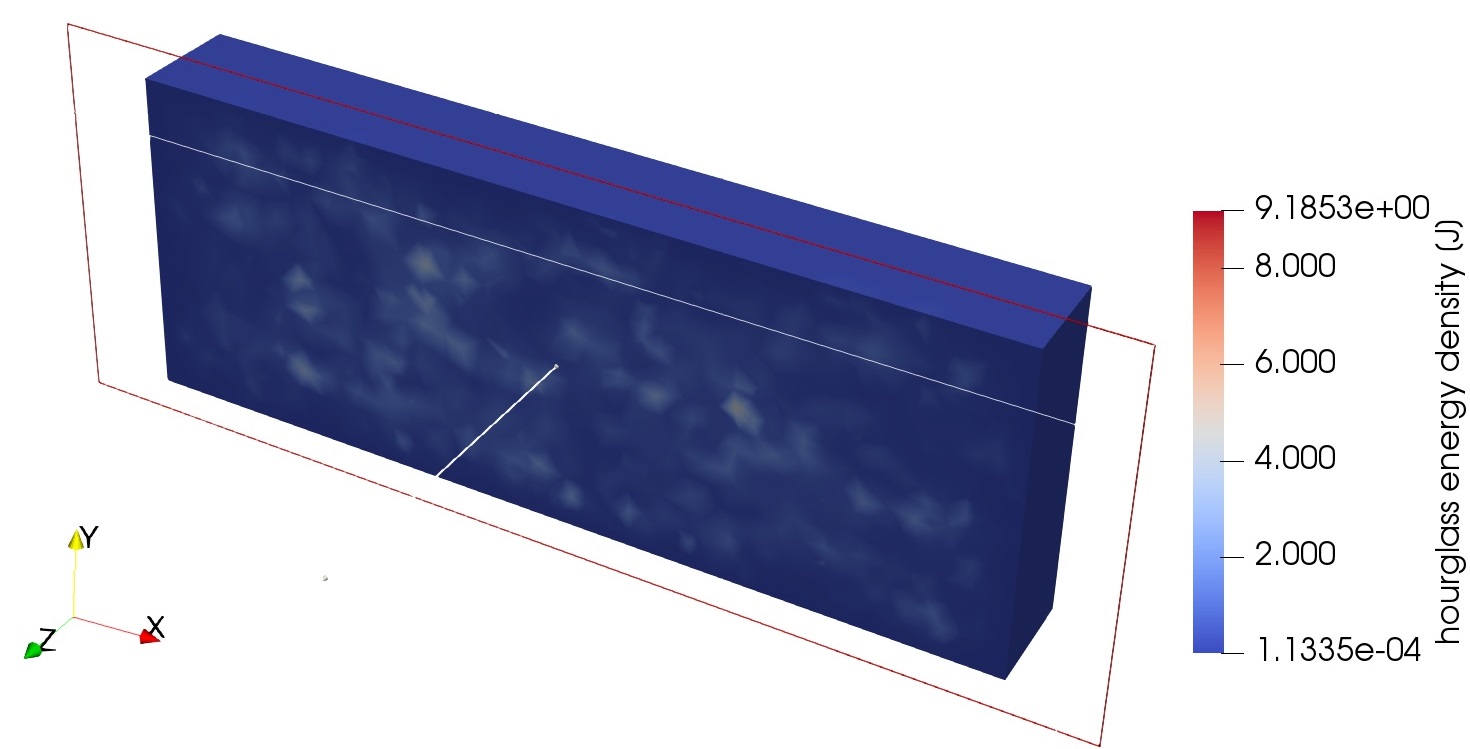}}
 \vspace{.1in}
 \subfigure[Finer mesh with hourglass control]{
 \label{fig:hgDensity4}
 \includegraphics[width=.4\textwidth]{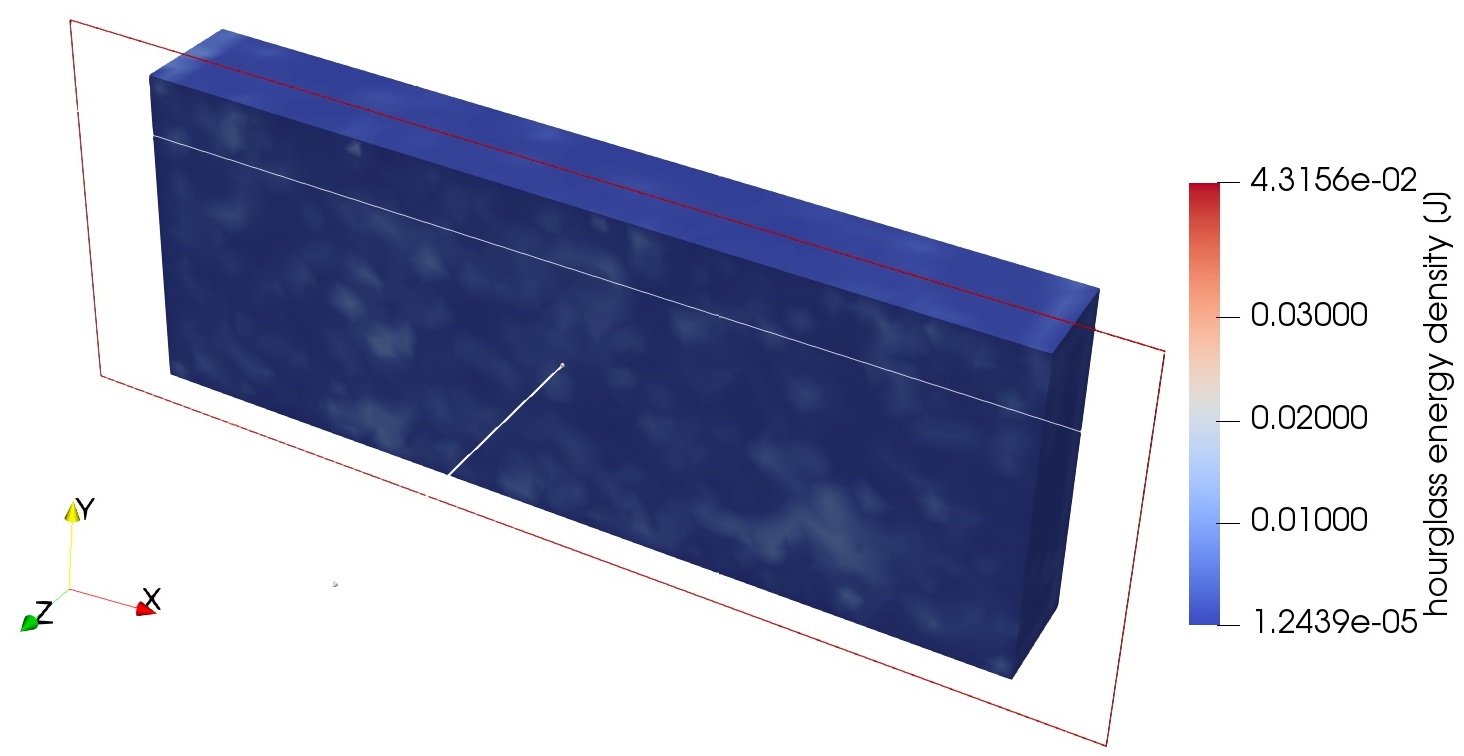}}\\
 \vspace{.1in}
\caption{Hourglass energy density on the clip of $z=1$ m.}
\label{fig:hgDensity}
\end{figure}

A three-dimensional cantilever beam loaded at the end with pure tension or compression of $P_x=1.0\times 10^6$ Pa is considered to test the performance of hourglass control. The dimensions and material parameters of the beam are the same as that in \S \ref{sbsec:3Dbeam}, as shown in Fig.\ref{fig:thickBeam}. The theoretical maximal displacement in $x$-direction is $(u_x)_{max}=2.6667\times 10^{-4}$ m. The total strain energy is $E_{strain}=800$ J. The particles on the left $yz-$plane are fixed in $x$ direction except one particle in $(0,0,0)$is fixed in all directions. Two discretizations the same as (a) Case 3 and (b) Case 4 in Table \ref{tab:beamStatistic} with/without hourglass control are tested. The $x$ displacement and hourglass energy density on the clip of $z=1$ m are shown in Fig.\ref{fig:hgUx} and Fig.\ref{fig:hgDensity}, respectively. The total strain energy and hourglass energy are given in Table \ref{tab:hgEnergy}. It can be seen that the hourglass control has significant influence on the accuracy of the solution.

\begin{table}[h]
\begin{center}
\begin{tabular}{ | c|c | c | c | c|}
\hline
 & (a) without HG & (a) with HG& (b) without HG& (b) with HG \\ \hline
		$E_{strain}^{numerical} $&874.92 &804.7&901.46&800.1\\ \hline
		$E_{hourglass}^{numerical} $& 6.442&0.0977&16.454&0.1051\\ \hline
	${u_x}_{max} $& 3.096E-04 & 2.849E-04 &3.147E-04 & 2.752E-04\\ \hline
\end{tabular}
\caption{Energy and maximal displacement in $x$ direction for 4 cases. HG denotes hourglass control.}\label{tab:hgEnergy}
\end{center}
\end{table}

\begin{figure}
	\centering
		\includegraphics[width=5cm]{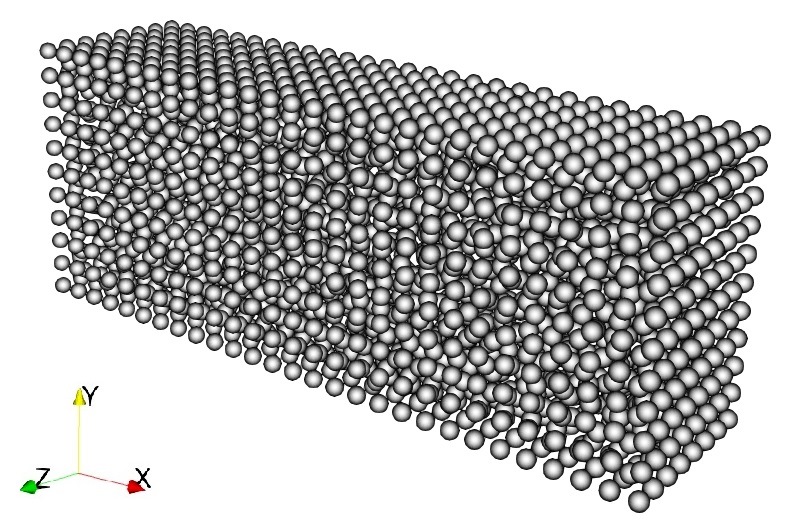}
	\caption{Irregular distribution of particles.}
	\label{fig:thickBeamParticle}
\end{figure}
\begin{figure}
 \centering
 \subfigure[]{
 \label{fig:bstrainXhg}
 \includegraphics[width=.4\textwidth]{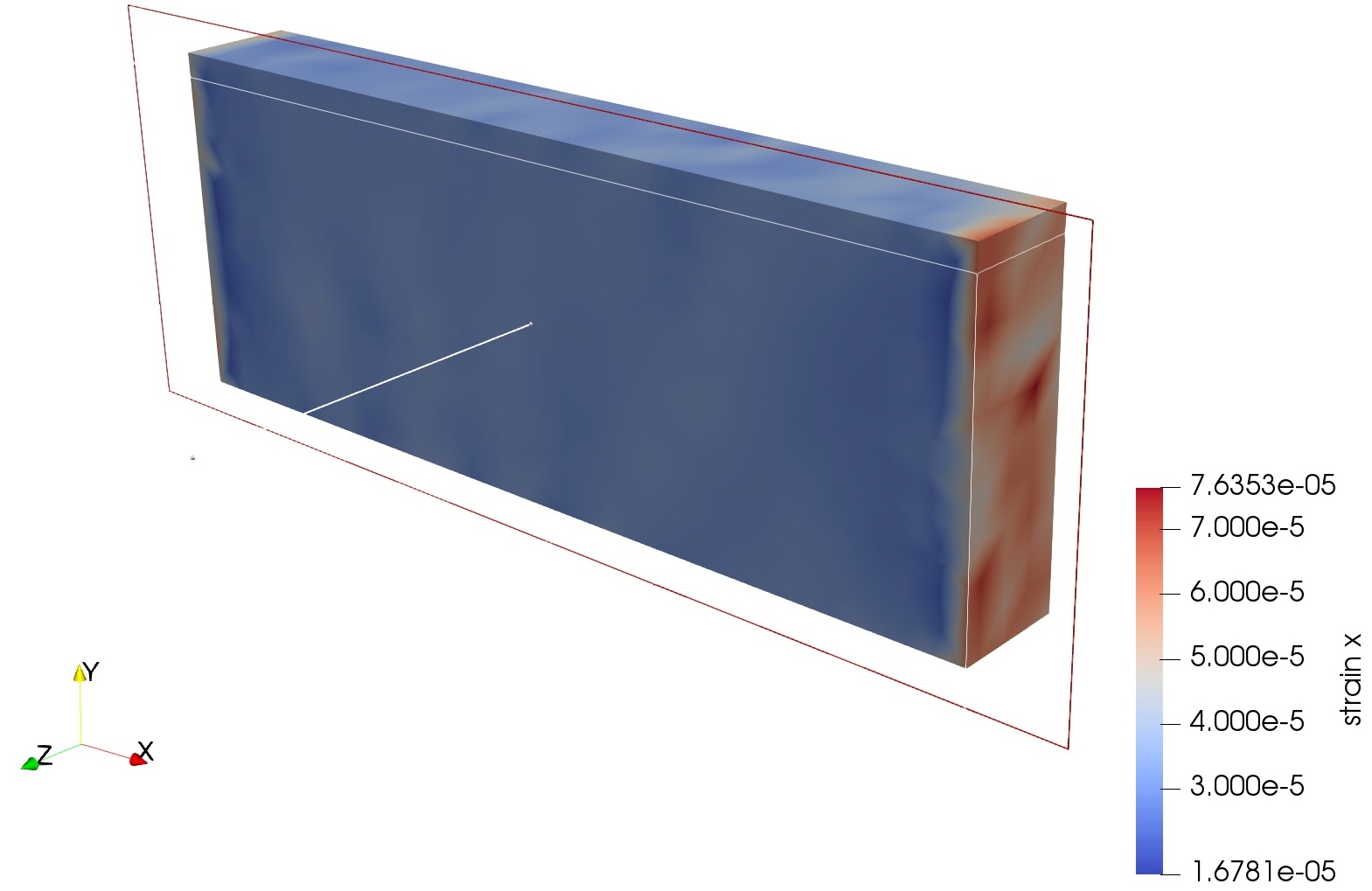}}
 \vspace{.1in}
 \subfigure[]{
 \label{fig:bstrainX}
 \includegraphics[width=.4\textwidth]{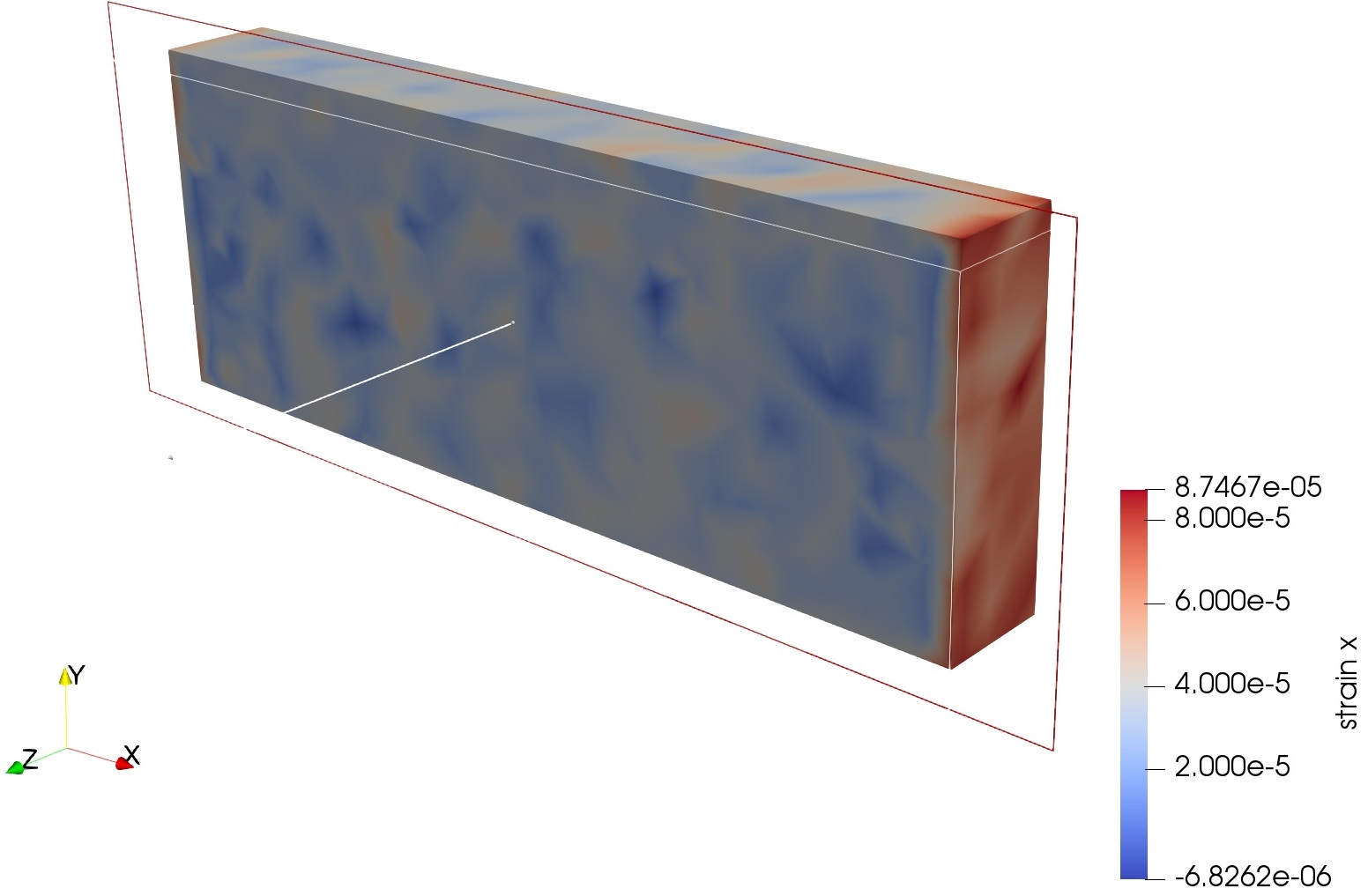}}
 \vspace{.1in}
\caption{Distribution of $x$ component of strain tensor, (a) with hourglass control; (b) without hourglass control.}
\label{fig:bsxStrain}
\end{figure}
\begin{figure}
 \centering
 \subfigure[]{
 \label{fig:bstrainEnerhg}
 \includegraphics[width=.4\textwidth]{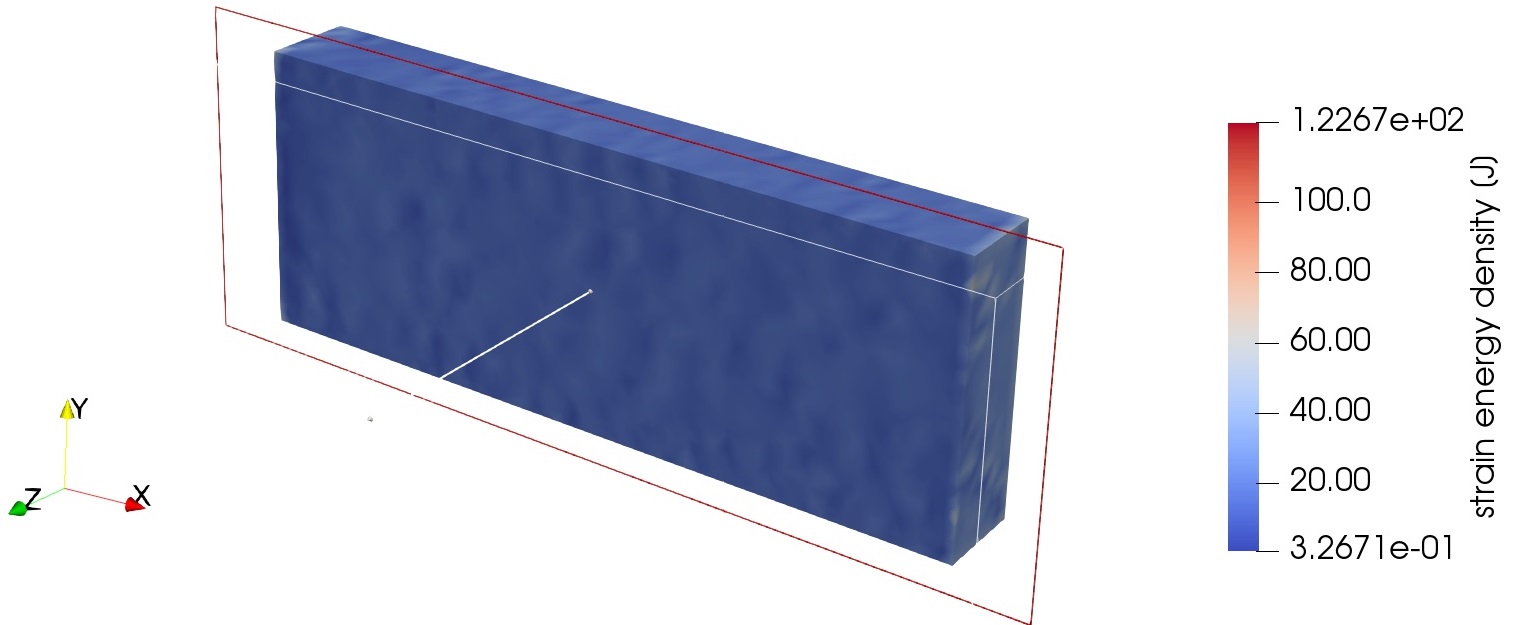}}
 \vspace{.1in}
 \subfigure[]{
 \label{fig:bstrainEner}
 \includegraphics[width=.4\textwidth]{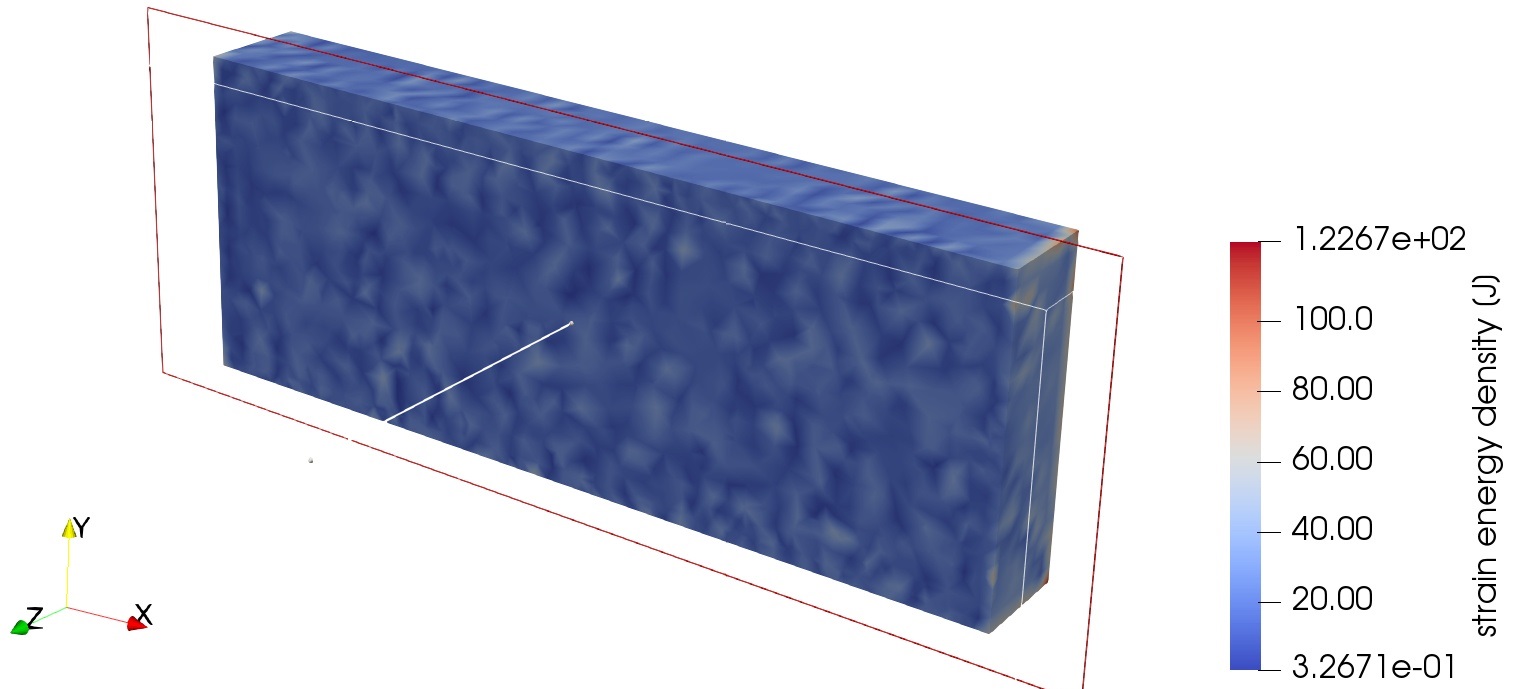}}
 \vspace{.1in}
\caption{Distribution of strain energy density, (a) with hourglass control; (b) without hourglass control.}
\label{fig:bsxStrainEnergy}
\end{figure}

Fig.\ref{fig:hgUx1} and Fig.\ref{fig:hgUx2} show that the hourglass control can effectively improve the result. For the pure tension test, the strain energy density and strain component in $x$ are evenly distributed for hourglass control, as shown in Fig.\ref{fig:bsxStrain} and Fig.\ref{fig:bsxStrainEnergy}.

The hourglass mode is obvious on the boundaries where the Dirichlet boundary and Neumann boundary are applied. The reason is that the boundary conditions are applied on only one layer of particles and the delta property is not well satisfied. 

The comparison in Fig.\ref{fig:bsxStrain} shows that hourglass control can effectively eliminate the hourglass mode, and make the strain field more smooth.

\subsection{Influence of smoothing length}
\begin{figure}
 \centering
 \subfigure[]{
 \label{fig:beam2Dmesh}
 \includegraphics[width=.6\textwidth]{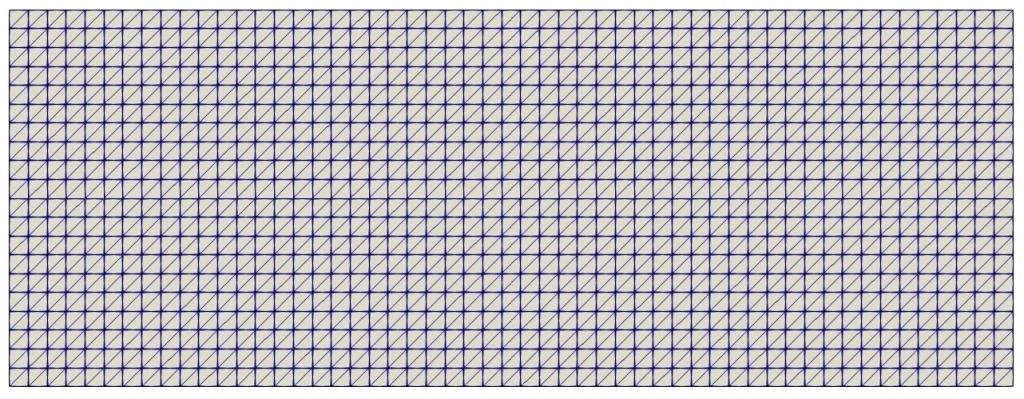}}
 \vspace{.1in}
 \subfigure[]{
 \label{fig:element2Node}
 \includegraphics[width=.25\textwidth]{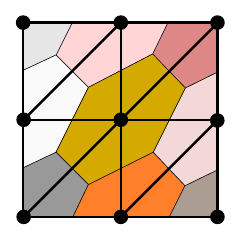}}
 \vspace{.1in}
\caption{(a) Mesh of 2D beam; (b) attach the element volume to nodes by average.}
\label{fig:beam2Dmesh2}
\end{figure}

One disadvantage of the implicit formula is {the high cost in assembling global stiffness matrix due to the large matrix sizes for each node}. In this section, we test the effect of smoothing length for quintic kernel function on the numerical accuracy in 2D plane stress solid. A thick beam in 2D with the same material parameters and dimensions in \S \ref{sbsec:3Dbeam} is considered. The hourglass energy control is used in all numerical examples of this section. The particles are constructed from the element given in Fig.\ref{fig:beam2Dmesh} by method shown in Fig.\ref{fig:element2Node}. The particle radius is estimated by the shape of disk. The smoothing length is selected as
\begin{align}
h_i=n \Delta x_i, \forall \bx_i \in \Omega\label{eq:hn}
\end{align} 

\begin{figure}
	\centering
		\includegraphics[width=9cm]{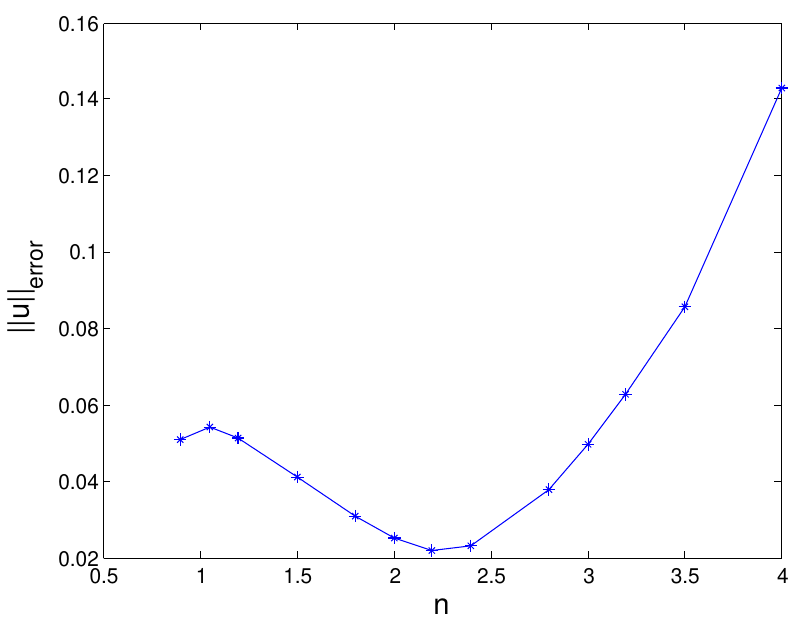}
	\caption{Displacement error for smoothing length.}
	\label{fig:hunorm}
\end{figure}
\begin{figure}
	\centering
		\includegraphics[width=9cm]{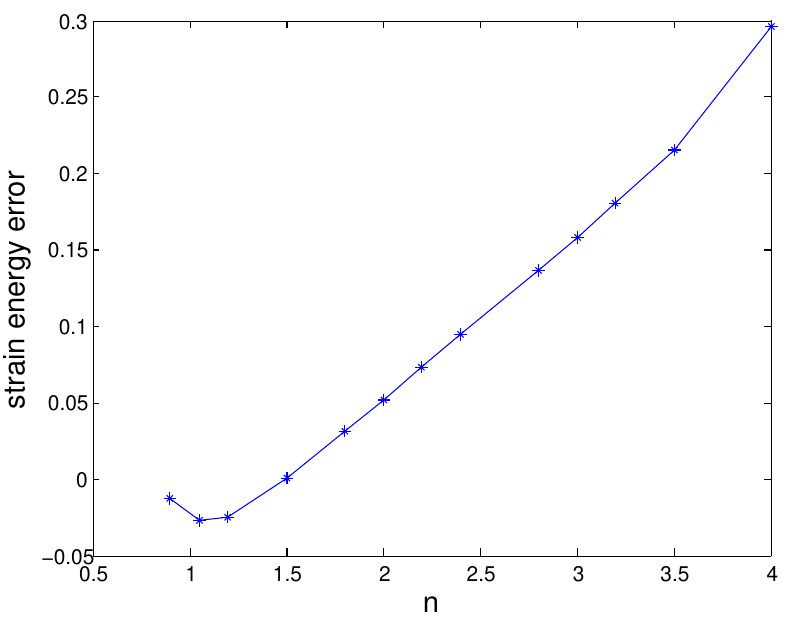}
	\caption{Strain energy error for smoothing length.}
	\label{fig:hEerror}
\end{figure}
\begin{figure}
	\centering
		\includegraphics[width=9cm]{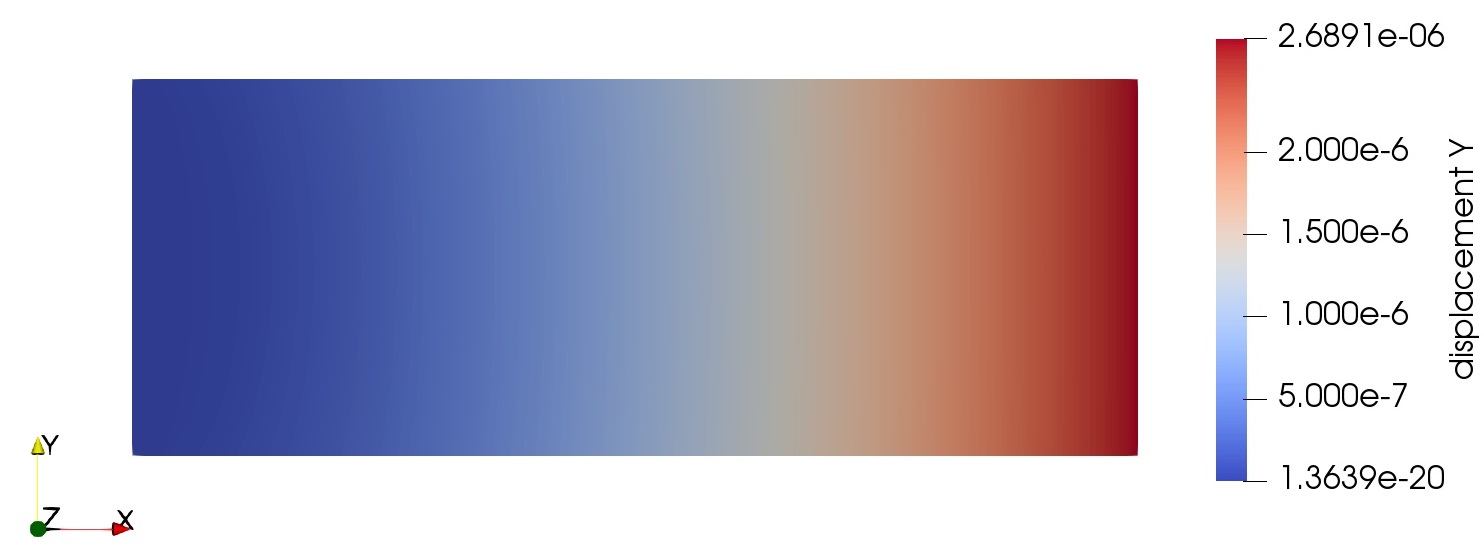}
	\caption{Displacement field for $n=0.9$ in Eq.\ref{eq:hn}.}
	\label{fig:u09}
\end{figure}
\begin{figure}
	\centering
		\includegraphics[width=9cm]{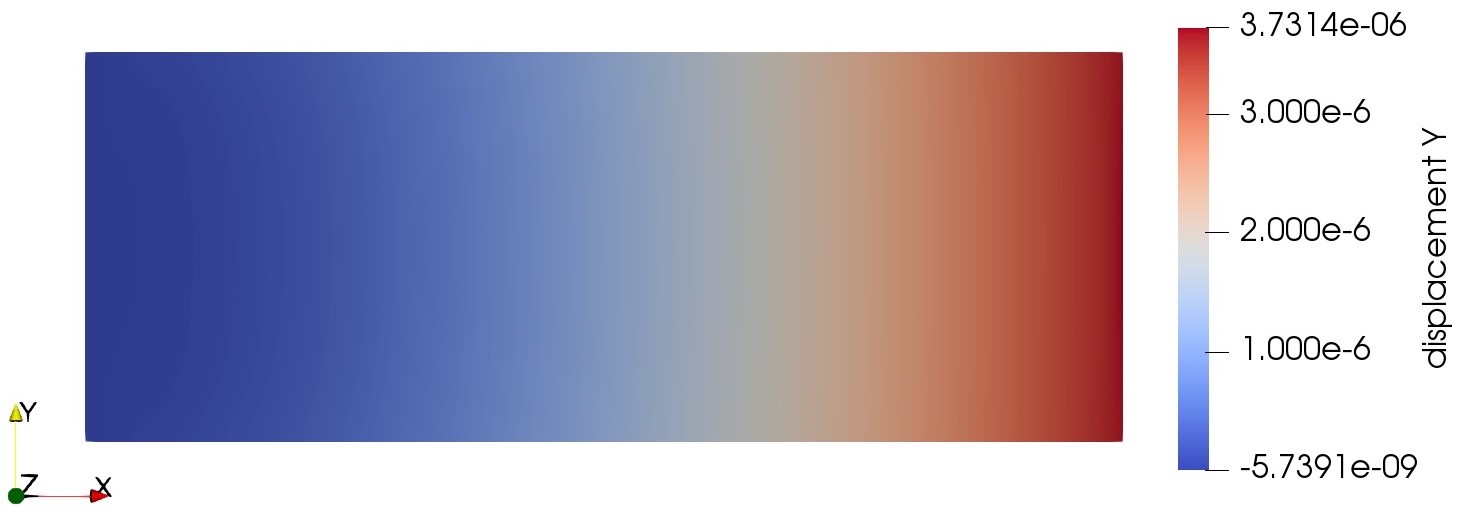}
	\caption{Displacement field for $n=3.8$ in Eq.\ref{eq:hn}.}
	\label{fig:u38}
\end{figure}
\begin{figure}
 \centering
 \subfigure[Case 1]{
 \label{fig:rubberPull1}
 \includegraphics[width=.3\textwidth]{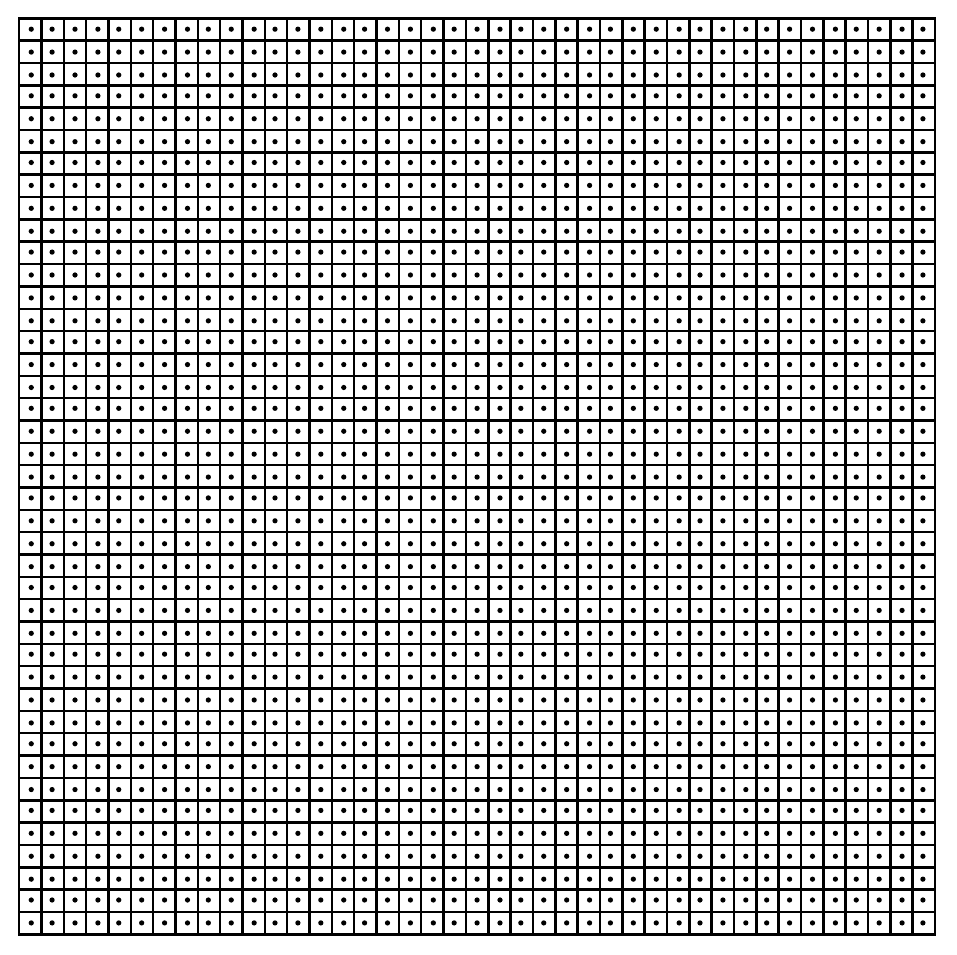}}
 \vspace{.1in}
 \subfigure[Case 2]{
 \label{fig:rubberPull2}
 \includegraphics[width=.3\textwidth]{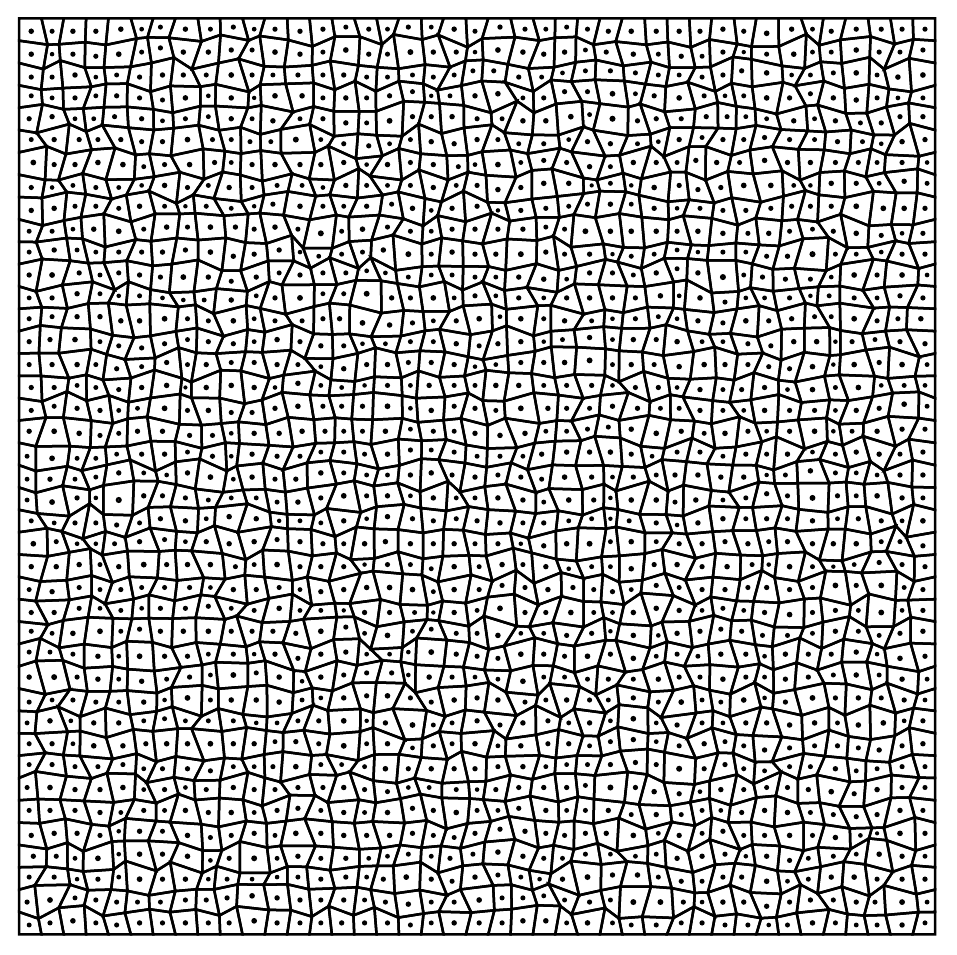}}\\
\caption{Discretizations of the rubber.}
\label{fig:rubberPull}
\end{figure}
\begin{figure}
 \centering
 \subfigure[$\alpha=0$ for Case 1]{
 \label{fig:rubberPullU1}
 \includegraphics[width=.3\textwidth]{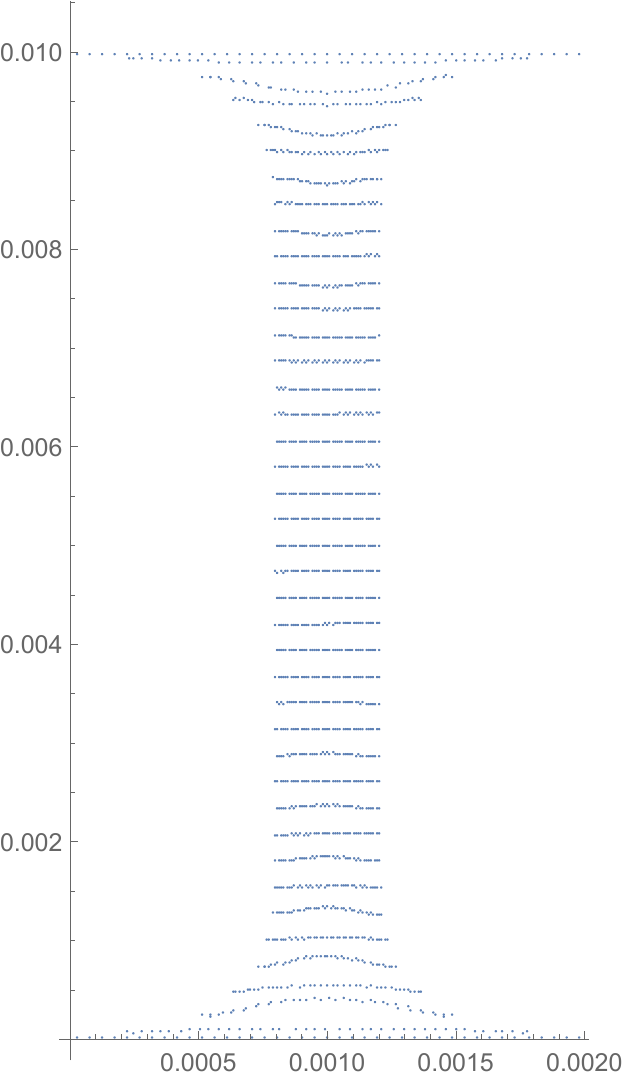}}
 \vspace{.1in}
 \subfigure[$\alpha=5 \mu$ for Case 1]{
 \label{fig:rubberPullU2}
 \includegraphics[width=.3\textwidth]{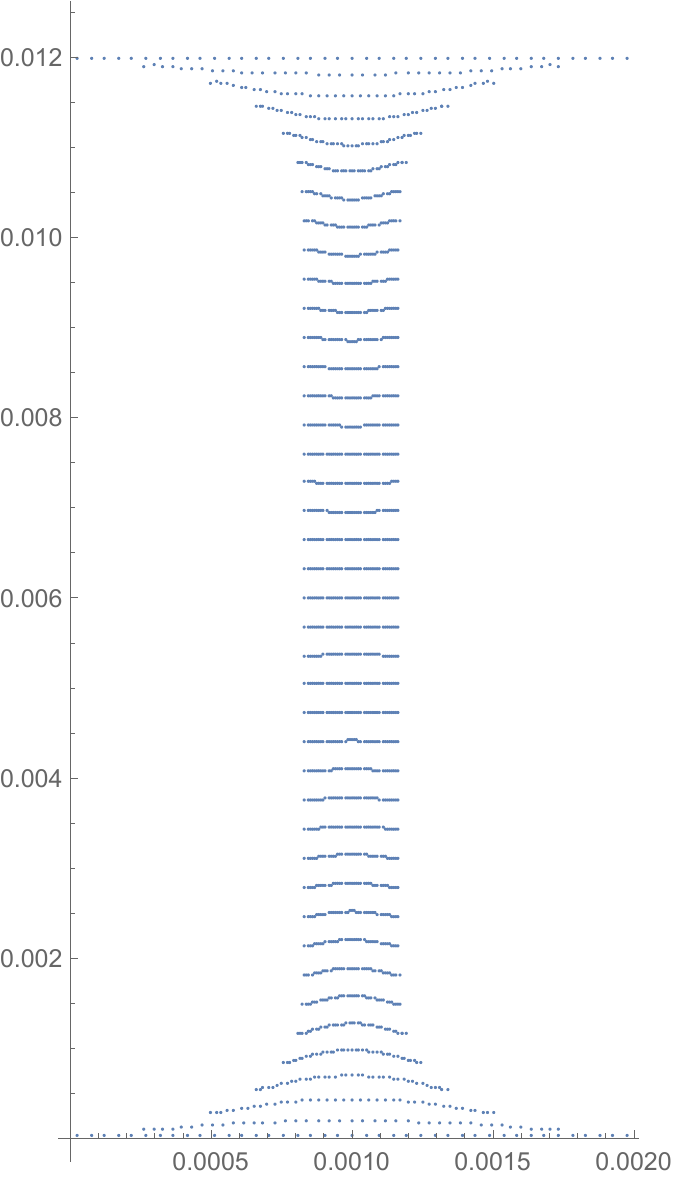}}\\
 \vspace{.1in}
 \subfigure[$\alpha=0$ for Case 2]{
 \label{fig:rubberPullU3}
 \includegraphics[width=.3\textwidth]{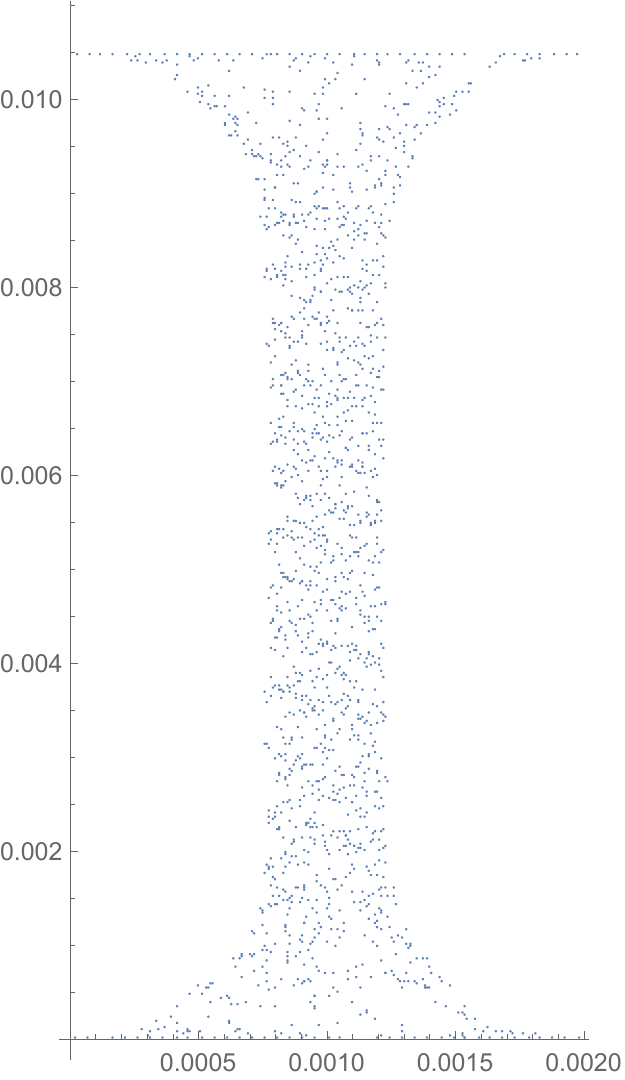}}
 \vspace{.1in}
 \subfigure[$\alpha=5 \mu$ for Case 2]{
 \label{fig:rubberPullU4}
 \includegraphics[width=.3\textwidth]{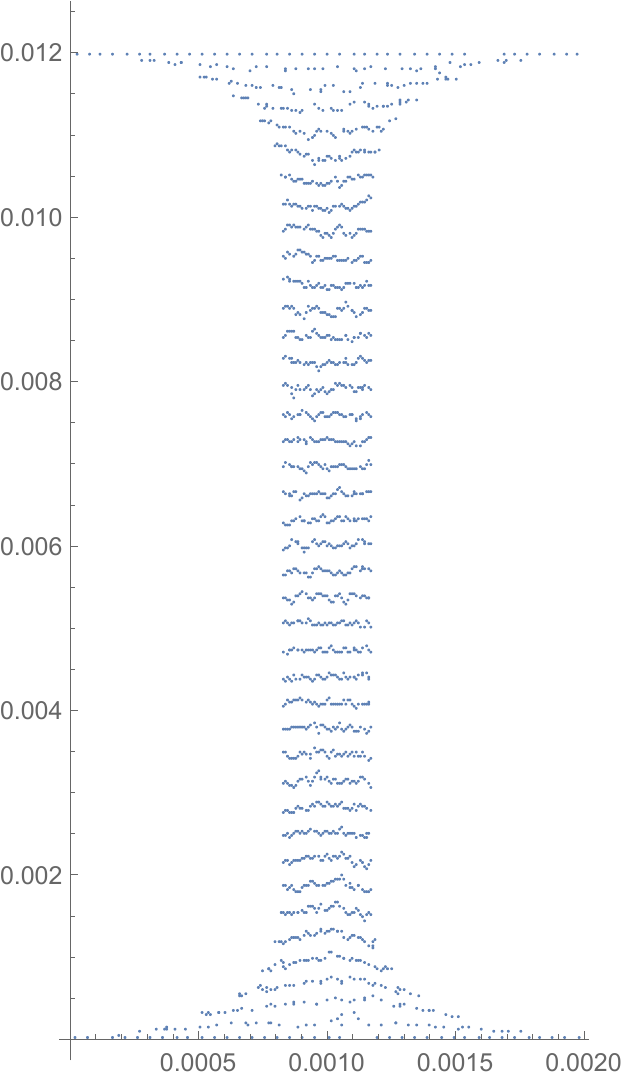}}\\
\caption{Final configurations of the rubber.}
\label{fig:rubberPullU}
\end{figure}

The $u_{error}$ and strain energy error are given in Fig.\ref{fig:hunorm} and Fig.\ref{fig:hEerror}. The number of neighbors for different smoothing length is given in Table.\ref{tab:neigh}. For the case of $n=0.9$, the minimal dimensions of the nodal stiffness matrix are 6x6, while the maximal dimensions of nodal stiffness matrix for case $n=3.8$ are 482x482. However, the larger smoothing length doesn't indicate a better numerical result. The ``optimal'' smoothing length scale for the corresponding kernel function is 2.2. When $n>2.2$, the numerical error increases with the smoothing length. On the other hand, the smoothing length scale $n=0.9$ offers good accuracy at the lowest computational cost. The displacement field for \hl{$n=0.9$ and $n=3.8$} are given in Fig.\ref{fig:u09} and Fig.\ref{fig:u38}, respectively.
\begin{table}[h]
\begin{center}
\begin{tabular}{ | c|c | c | c | c| c| c|c|}
\hline
smoothing length & 0.9 & 1.2 & 1.5 & 2 & 2.4 & 3 & 3.8 \\ \hline
		min&2 &3&3&7&10&14&21\\ \hline
		max&12 &20&36&68&96&146&240\\ \hline
\end{tabular}
\caption{Number of neighbors}\label{tab:neigh}
\end{center}
\end{table}
\hl{\subsection{Rubber pull test}\label{sec:rubberPull}
In this section, we test a rubber with 500\% elongation based on hyperelastic material in Eq.\ref{eq:neohook2}. The initial dimensions of the plate are $[0,2]\times [0,2]$ mm$^2$. The material parameters are elastic modulus $E=0.2$ MPa and Poisson ratio $\nu=0.45$. Two discretizations with regular particle distribution and irregular particle distribution are tested, as shown in Fig.\ref{fig:rubberPull}. The area of each particle is selected as that of the element. Hourglass penalty with $\alpha=0$ and $\alpha=5 \mu$ are tested, where $\mu=\tfrac{E}{2(1+\nu)}$. The final deformation of four cases are shown in Fig.\ref{fig:rubberPullU}. The smoothing length for each particle is selected as the maximal distance with respected to its 12 nearest neighbors. In order to reach the specified elongation, the up layer of the particles are displaced gradually by 10 mm in several increments. The Newton-Raphson iteration algorithm is adopted to solve the equations. For case 1, the simulation without hourglass control does not converge when the displacement on the boundary is larger than 8 mm, as shown in Fig.\ref{fig:rubberPullU1}. In the numerical simulation, the hourglass control stabilizes the scheme and makes the convergence easier. For Case 2, the model with irregular particle distribution and without hourglass control becomes unstable and diverges when the displacement on the boundary is larger than 8.25 mm, as shown in Fig.\ref{fig:rubberPullU3}. When hourglass control is applied, the modeling based on irregular particle distribution is very stable throughout the simulation.}

\subsection{Large deformation problem}
\begin{figure}
	\centering
		\includegraphics[width=6cm]{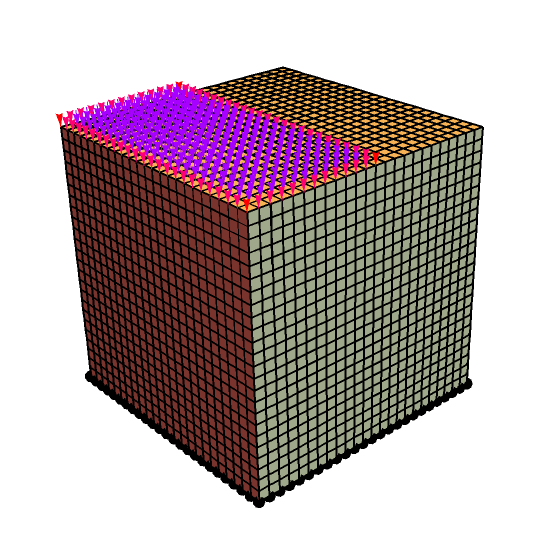}
	\caption{Boundary condition of the cube, where the upper surface is applied with $p_z=-2 N/m^2$ and the nodes on the bottom are fixed in all direction.}
	\label{fig:ldbc}
\end{figure}

\begin{figure}
 \centering
 \subfigure[]{
   \label{fig:lddeform1}
   \includegraphics[width=.4\textwidth]{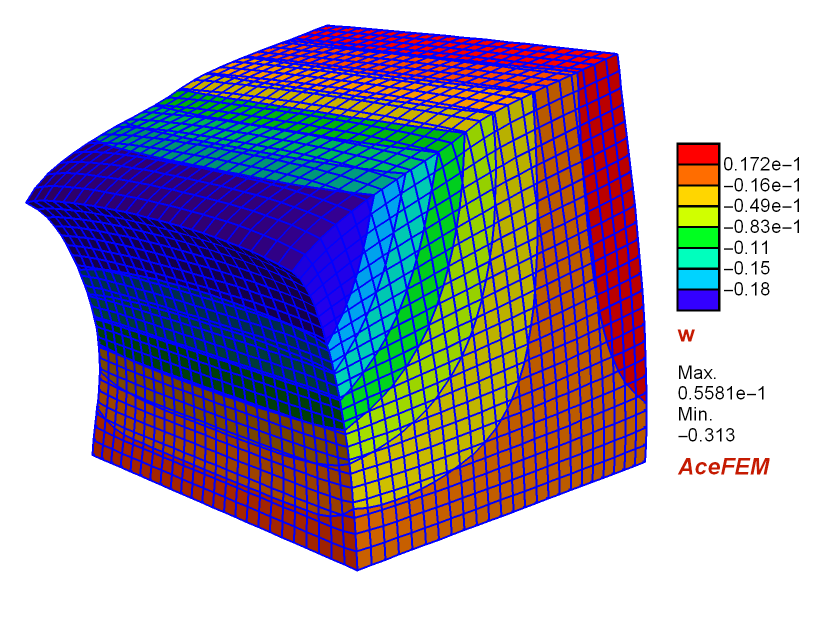}}
 \vspace{.1in}
 \subfigure[]{
   \label{fig:lddeform2}
   \includegraphics[width=.4\textwidth]{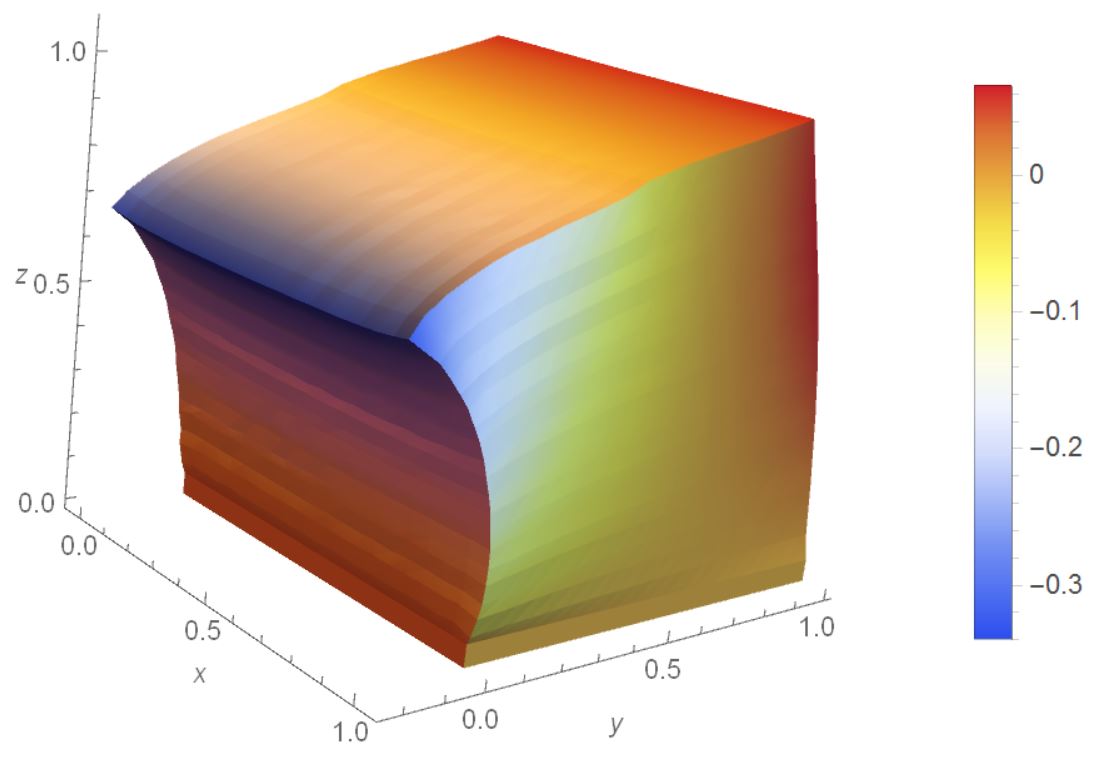}}
\caption{(a) Final deformation by finite element method. (b) Final deformation by implicit SPH.}
\label{fig:lddeform}
\end{figure}

\begin{figure}
 \centering
 \subfigure[]{
   \label{fig:ldwv1}
   \includegraphics[width=.4\textwidth]{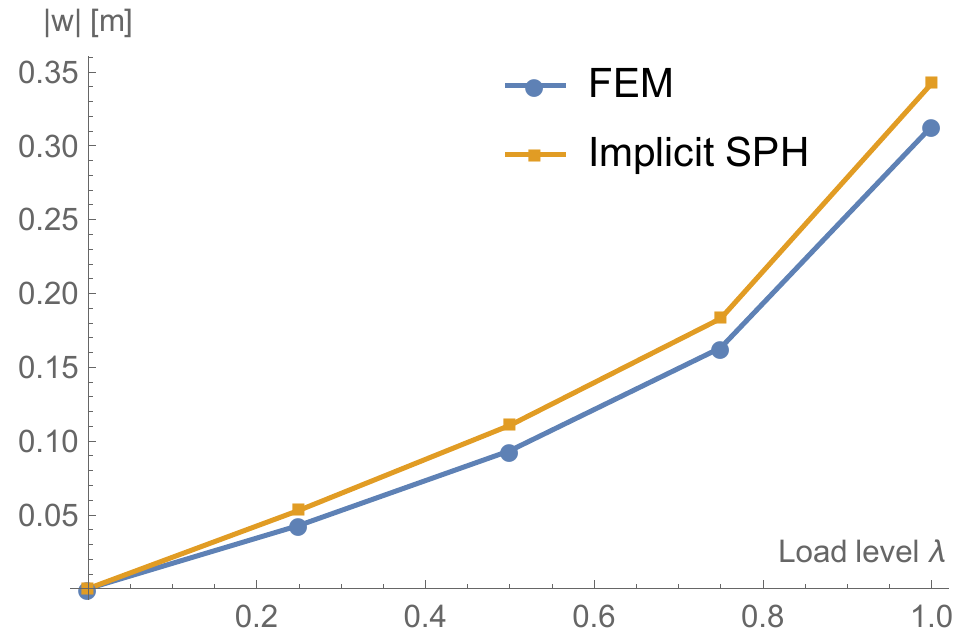}}
 \vspace{.1in}
 \subfigure[]{
   \label{fig:ldwv2}
   \includegraphics[width=.4\textwidth]{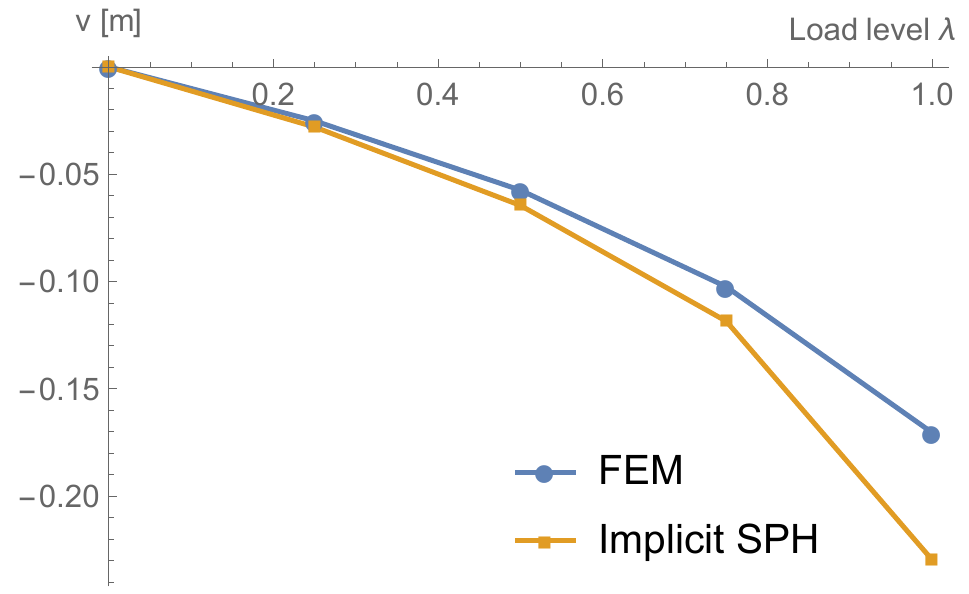}}
\caption{Displacement of point ($0,0,L$) vs load levels. (a) Displacement in $z$-direction. (b) Displacement in $y$-direction.}
\label{fig:ldwv}
\end{figure}

In this section, a cube of length $a=1$ m with hyperelastic material given by Eq.\ref{eq:nhm} is modeled. This example demonstrates the capability of current formulation in solving problems involving geometric nonlinearity. The material parameters for the cube are elastic modulus $E=10 N/m^2$, Poisson ratio $\nu=0.3$. The nodes on the plane $z=0$ are fixed in all directions and the surface $\{x,y,z\}\in [0,a]\times [0,a/2] \times a$ are applied with pressure load $p_z=-2 N/m^2$, as shown in Fig.\ref{fig:ldbc}. The cube is discretized into $21^3$ particles. The load is applied on 4 sub-steps and in each sub-step the Newton-Raphson iteration is used to find the equilibrium, where the convergence is reached when the residual norm is less than $10^{-6}$. The quintic kernel function is used and smoothing length is $h_i=2.05 \Delta x_i$. The numbers of iteration for four sub-steps are (4,4,5,7) sequentially. The nonlinear effect increases with the load levels and more iteration is required to achieve the convergence. The final deformed configuration for implicit SPH and implicit FEM are given in Fig.\ref{fig:lddeform}, where the deformation is quite similar. The FEM result is implemented in the AceFEM environment \cite{korelc2006acegen,korelc2016automation}. The largest displacement in $z$-direction are (-0.313 m,-0.342 m) for FEM and implicit SPH, respectively. The displacement in $y,z$-direction for different load level is depicted in Fig.\ref{fig:ldwv}. The difference for the maximal deformation in $z$-direction is approximately $9.2\%$, which is due to that SPH being a meshless method does not possess the Kronecker-delta property.

\vspace{-6pt}
\section{Conclusions}
In this paper, we derived the dual-support SPH by means of variational principle and demonstrated that the implicit form of SPH can be obtained with ease. During the evaluation of nodal stiffness matrix, only the variation of deformation gradient is required. We also show that the hourglass control is necessary to in the SPH solid. We presented a general framework for the implicit SPH analysis which allows for material nonlinearity and geometrical nonlinearity. The fluid version of dual-support SPH is presented in the other paper. \hl{The proposed implicit SPH formulation obtains the residual and tangent stiffness matrix in a way quite similar to the finite element method. Many problems solved by FEM can be solved by the current scheme with some adaption. For examples, implicit SPH can replace the finite element formulation in the phase field fracture method \cite{miehe2010thermodynamically,zhou2018phase2,zhou2018phase3} to solve the fracture problems}.
\vspace{-6pt}
\section*{Acknowledgments}
The authors acknowledge the supports from the RISE-BESTOFRAC, COMBAT Program (Computational Modeling and Design of Lithium-ion Batteries, Grant No.615132).
\vspace{-6pt}
\section*{References}
\bibliographystyle{unsrt}
\bibliography{sphimplicit/SPHImplicit}
\end{document}